\documentclass[12pt]{elsarticle}

\usepackage[hidelinks]{hyperref}
\usepackage{adjustbox}
\usepackage{multirow}
\usepackage{lscape}

\journal{Computers \& Education}

\newcommand*\mean[1]{\bar{#1}}

\biboptions{authoryear}

\begin{document}

\begin{frontmatter}

\title{Network analyses of student engagement with online textbook problems}

\author{Jesper Bruun}
\address{Department of Science Education, University of Copenhagen, Øster Voldgade 3, Copenhagen K, DK-1350}

\author{Pia J. Ray}
\author{Linda Udby}
\address{X-ray and Neutron Science, Niels Bohr Institute, University of Copenhagen DK-2100}
\clearpage
\begin{abstract}
Problem solving in physics and mathematics have been characterized in terms of five phases by Schonfeld and these have previously been used to describe also online and blended behavior. We argue that expanding the use of server logs to make detailed categorizations of student actions can help increase knowledge about how students solve problems. 
We present a novel approach for analyzing server logs that relies on network analysis and principal component analysis. We use the approach to analyze student interactions with an online textbook that features physics problems. We find five 'components of behavioral structure': Complexity, Linear Length, Navigation, Mutuality, and Erraticism. Further, we find that problem solving sessions can be divided into three over-arching groups that differ in their Complexity and further into ten clusters that also differ on the other components. Analyzing typical sessions in each cluster, we find ten different behavioral structures, which we describe in terms of Schonfeld's phases. We suggest that further research integrates this approach with other methodological approaches to get a fuller picture of how learning strategies are employed by students in settings with online features.

\end{abstract}


\end{frontmatter}
\clearpage 

\section{Introduction}
\label{S:1}
Problem solving in physics and mathematics has been the object of a large amount of research over a long period of time \citep{larkin1980expert,schoenfeld1992,chi1989self,hsu2004resource,docktor2014synthesis}. Specifically, work has been done to utilize digital technologies to support student autonomy while solving problems \citep{pol2005solving,harskamp2007schoenfeld,pol2008effect,pol2009,hsu2004resource}. With web-based technologies, student actions can be recorded in \emph{server logs} \citep{romero2008data}. When students engage with online teaching material, server logs capture detailed information including where students click, how they scroll, what input they make in text-fields, where they drag items. We argue that using server logs to make detailed categorizations of student online actions can provide unique knowledge about student problem solving and the affordances of online problem solving.  

Solving end-of-chapter text-book problems are sometimes seen as part of becoming a scientist \citep{reif1999teaching}. These kinds of problems have been investigated for decades, and this has resulted in both fundamental insights into how students categorize problems, the strategies they employ to solve them, and how they understand different representations involved in problem solving \citep{docktor2014synthesis}. Students have traditionally either been observed while problem solving, and their behavior then analyzed qualitatively \cite[see e.g.][]{chi1989self}, and/or their performance after exposure to different treatments evaluated \cite[see e.g.][]{pol2008effect}. Only recently have researchers started using data mining of student choices to inform teaching and learning \citep{ferguson2012learning}. In this article, we utilize the functionality of web-based text in a novel way: online presentations of textbook material and end-of-chapter problems with an option to show or hide hints and solutions affords detailed analyses of the traces students leave behind when solving problems. By embedding problems within an online textbook format, it is possible both to track (1) what students do online when directly engaging with the problem text, hints, and solutions and (2) how they access textbook material as part of the process. This has the potential to yield insights in to the many ways in which students may engage with problem solving and thus to aid instruction. 

As a case, this article investigates end-of-chapter-like problems in an online, wiki-based textbook environment in the Physics discipline of Neutron Scattering. The aim is to create an empirically based typology of online end-of-chapter problem solving actions for this environment that is rooted in the literature concerning problem solving. To create such a typology, we propose to extract records of student interactions with a web-based environment (we call these records \emph{sessions}) from server logs and use network analysis to find clusters of similar records. Although the present work is embedded in a "wiki-textbook" format, the method of analysis is general. It requires only that actions of students or student groups be put into a meaningful ordered sequence. 
  
This article proceeds in Section~\ref{sec:litRev} with background on online problem solving and network analysis of logs of student online actions.  Section~\ref{sec:RQ} clarifies the terminology we use regarding student online actions and student behaviors, and then states three research questions to be answered. Section~\ref{sec:wikibook} describes the teaching material; specifically the types of problems students have been solving. Section~\ref{sec:methods} provides an overview of the proposed methodology, the details of which can be found in ~\ref{app:detailsMethodology}. Section~\ref{sec:results} shows the results and present our interpretation of the results. Finally, Section~\ref{sec:discussion} discusses the results in light of (1) student behaviors while solving problems, and (3) limitations of the study. 

\section{Background}
\label{sec:litRev}
\subsection{Students' use of hints and solutions in an online setting}
Using hints and worked out solutions has been advocated in the literature as means to increase students' problem solving skills \citep{sandelin2011,pol2005solving,pol2008effect,pol2009,harskamp2007schoenfeld}. 

In describing problem solving, \citet{schoenfeld1992} identifies five cyclical episodes/phases, which students go through when solving mathematical problems. These phases can be summarized as (1) surveying the problem (read, analyze), (2) activating knowledge (explore), (3) making a plan (plan), (4) carrying out the plan (implement), and checking the answer (verify) \citep{harskamp2007schoenfeld}. While the theory has been illustrated by a number of authors, the full theoretical framework is too elaborate for the purposes of this paper. However, it is interesting to note that actively surveying the problem, activating knowledge, and checking the results afterwards are linked to fruitful problem solving behavior. 

While seemingly linear, the process likely involves jumps between different phases, in that students may, for example, interrupt making a plan in order to re-read the problem \citep{wilson1993mathematical}. \citeauthor{schoenfeld1992} gives detailed examples of how this plays out in mathematics. \citet{harskamp2007schoenfeld} apply Schoenfeld's model to create a controlled learning environment, where students use hints to facilitate different parts of the problem solving process. The environment features specific hints for the read and analyze phase, for the exploration phase, and for the planning phase. Furthermore, students have the opportunity to verify their solutions. Using data on student use of hints as input, \citeauthor{harskamp2007schoenfeld} construct a structural equation model that shows significant links between both using hints and verifying solutions on post-test scores. Also, their model suggests that an important part of the problem solving process is the first phase of reading and analyzing the problem. 

As compared with a worked out solution in a traditional textbook, being able to hide and show a solution may afford different behaviors. For example, it may be easier to look quickly at part of the solution and then hide to see if one can use that part to solve the whole problem. \citet{pol2005solving} find that the use of digital hints and solutions outperformed a traditional textbook problem solving class quite substantially with an effect size of 0.89. They were further able to show that students that had used digital hints and solutions were better at analyzing the problem and planning their solutions. 

Timing the availability of hints and solutions may be important in some contexts. For example, \citet{pol2008effect} show that providing hints during a session and worked examples after a session was more effective than only providing solutions after a session, attributing the effect to a combination of to practicing and systematic use of hints. Interestingly, other research suggests practicing is not an important parameter when learning how to solve problems in physics \citep{kim2002students} and that transfer to new types of problems is difficult \cite{mestre2006transfer}. Thus, it seems more likely that the effect of hints and worked examples is coupled to the strategies students employ in order to become better at problem solving. 

Common to studies found in the literature is that controlled environments have been created so that students' choices are limited. Such environments will guide students towards a particular behavior. While this may be suitable for students in some contexts, master level university courses in the sciences often involve demands for self-regulated learning \citep{zimmerman1989social,broadbent2015self}. Self-regulated learning can be seen as a triadic relationship between processes of meta-cognition, observable behaviors, and changing the environment to fit one's needs. Self-regulation is then facilitated by self-observation, self-judgment, and self-reaction. In terms of self-regulated learning, online digital textbooks with little or no guidance may serve as an arena for students to develop new strategies for linking their current thinking about a problem while using the course content. Not much is known about detailed student actions in such arenas. However, we propose that data-mining and specifically network analysis may be used to gain such knowledge. 


\subsection{Using network analysis for mining data logs}
Online textbooks affords detailed analyses of parts of students' interaction with course teaching material. Server logs provide rather detailed registers of the interactions between student and web-page. Server logs can be used to provide information about sequences of interactions with the web-page, and since sequences can be clustered in terms of their similarity \citep{xing2010sequence}, it follows that server logs can be used to cluster sequences of student actions. We propose to use network analysis to find clusters and describe our approach in Sectio~\ref{sec:methods}. 

A network is a collection of entities and connections between these entities \citep{bruun2016networks}. The context defines what the entities represent. In network terminology, these entities are called nodes and their connections are called links. Most work in educational settings has been centered on social network analysis \citep[see e.g.][]{macfadyen2010mining,garcia2014data}, but recently network analysis has also been employed in educational data mining to find connections between words used by students \citep{dascalu2014mining,rabbany2014collaborative}. For example, \citet{rabbany2014collaborative} combines text-mining techniques and word-co-occurrence to find clusters of topics with which students engage.

Studies have used network analysis to investigate students' behavior as observed in game-like teaching situations \citep{shaffer2009epistemic}, in interview settings \citep{bodin2012mapping}, and in classroom settings \citep{bruun2017network}. These networks depict the actions of a student, a group of students, or a classroom of students in a particular context and as coded by an observer. These \emph{action networks} \citep{bruun2016networks} can be constructed in different ways. \citet{shaffer2009epistemic} use co-occurrence of predefined codes in a period of time as a way to define nodes (codes) and links (co-occurrence). \citet{bodin2012mapping} and \citet{bruun2017network} also use predefined codes to classify different types of actions, but for them links are based on the time-order of occurrence. Thus, nodes represent codes and links represent the ordered sequence of two codes. These two approaches seem in line with the previously presented theoretical view on problem solving as involving stages that follow each other in ordered cycles.

The approaches above will produce an action network for each observed (or recorded) teaching and learning activity, for example, problem solving classes or discussions. This could result in potentially many networks, and since our aim is to find clusters of similar networks, a strategy is warranted. One strategy is proposed by \citet{faust2006comparing}, who uses correspondence analysis to find similarities between 52 social networks and then an approach reminiscent of $k$-means to find clusters. In this article, we will follow \citet{dolin2018SAD}, who use principal component analysis to find similarities between networks depicting student-teacher dialogue and subsequently created a similarity network in which clusters of dialogues could be found. This strategy can be seen to have two significant advantages over commonly used clustering techniques, such as $k$-means and hierarchical clustering \citep{dutt2015clustering}. First, the quality of the clustering can be assessed in terms of a measure called the modularity ($Q$), which is the fraction of connections within a cluster minus what could be randomly expected. For $Q<0.3$  \citep{newman2004fast}, there would be no significant community structure to detect, and most community detection methods search the solution with the highest $Q$ \citep{lancichinetti2009community}. Second, more information about the structure of a community is kept; it is not given that the clustering structure is hierarchical or flat, it could be either or in-between. Network community detecting thus provides means to investigate such structures in more detail. 

\section{Research Questions}
\label{sec:RQ}
In this study, we are using only server logs to find patterns in student actions. Thus, any relation to learning strategies or behaviors employed while learning must be inferred. Before proceeding, we make the distinction between learning behaviors, which is what the students do in their learning processes, and \emph{behavioral structures} as they appear in networks of student actions. We define behavioral structures as the trace left behind by online actions. Our analysis of server logs for this article is meant to create a typology of behavioral structures. Furthermore, the correspondence between behavioral structure and networks suggest that behavioral structures could be seen as constituted by basic components that make a more complex whole. With these distinctions between learning behavior, behavioral structures, and components of behavioral structure, our research questions are:

\begin{itemize}
\item RQ1: Which components of behavioral structure relevant to problem solving can be identified by analyzing action networks of server logs?
\item RQ2: Which types of action networks relevant to problem solving can be identified by using the identified components, and how can these be characterized?
\item RQ3: Which behavioral structures in regards to problem solving may be inferred from the resultant clusters session networks?
\end{itemize}

To answer the research questions, we propose a two-level network analytical approach. At one level, we create action networks that depict student online behaviors. At another level we create a network of similar action networks. The problems we are going use as a case for our analysis, are situated in a particular environment, which will necessarily affect the interpretations we can make. Therefore, in the next section, we describe this environment; the wiki-textbook. 

\section{The wiki-textbook}
\label{sec:wikibook}
Studies in educational research have investigated the use and potential of wikis as student collaborative platforms \citep{augar2004,parker2007,lin2009,matthew2009,karasavvidis2008}, whereas educational research on the use of teacher-produced wikis as teaching material for students seems to be lacking. The wiki-textbook utilizes the possibility of using the wiki-format to create a textbook edited by experts within a particular knowledge domain. This entails a constantly updated textbook, with the functionality of a wiki. Such a textbook does not entail collaboration between many disparate sources like Wikipedia. Rather, it is the focused enterprise of scientists with very specific knowledge. 

The wiki-textbook as used in this study is organized in a tree-structure that mirrors a standard textbook. Each wiki-page can be seen as a sub-section. Subsections are collected and shown in  section wiki pages. Section pages are collected and shown in chapter wiki-pages. All text chapters are listed in a menu at the main page of the wiki-textbook and problems are placed in separate sections. Each problem has a dedicated wiki-page, and problems are collected according to relevant chapters. The reason for this structure is to limit the need for scrolling \citep{waestlund2005,singer2017reading}, since text-passages become smaller for each subdivision. 

Problems in the wiki-textbook make use of hints and model solutions in the sense that they can be shown and hidden again by the user. Hints and solutions have been developed separately over roughly five years taking typical student procedural and content-related questions for solving each problem into account. Consistent with the finding that solving many of the same types of problem need not lead to better problem solving skills \citep{kim2002students}, each problem has to do with a particular concept or situation that is relevant to neutron scattering. Furthermore, users are students at the graduate level, which suggests that they likely will employ deep learning strategies when using hints and solutions \citep{sandelin2011}. Thus, unlike learning material employed in related research \citep{pol2005solving,harskamp2007schoenfeld,pol2008effect,pol2009}, the wiki-textbook does not focus on development of student problem solving strategies.

\begin{figure}
\label{fig:exampleProblem}
\includegraphics[width=\textwidth]{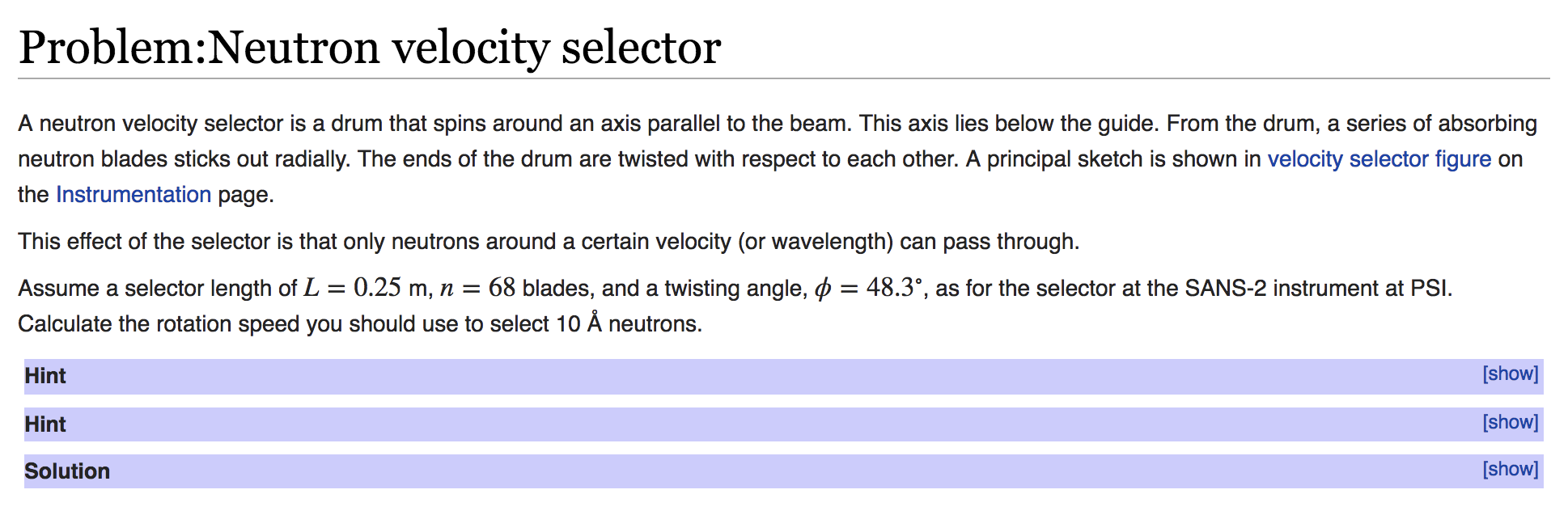}
\includegraphics[width=\textwidth]{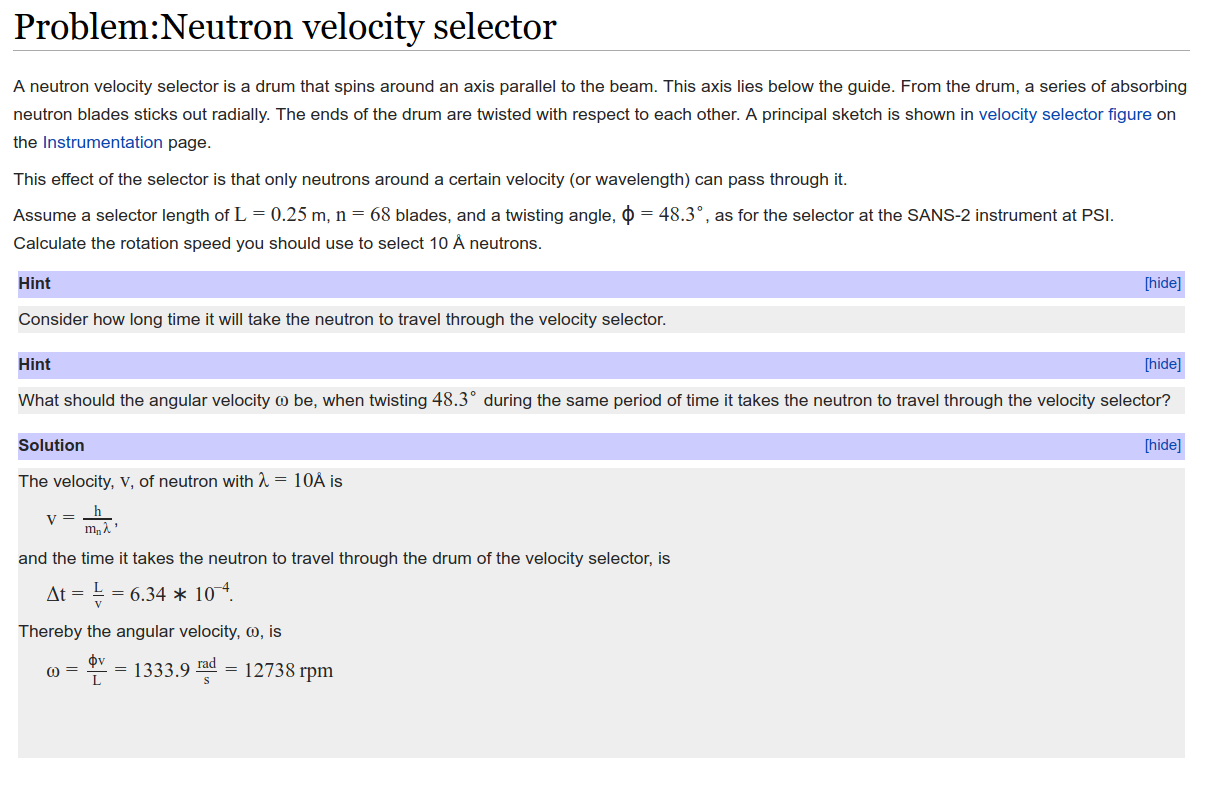}
\caption{An example of a wiki-textbook problem with no hints or model solution shown (top) and with all hints and the model solution shown (bottom).}
\end{figure}

Hints and solutions to problems are accessed by students at their own leisure thus enabling self-regulated differentiated teaching for students. This means that they decide individually when they need a hint to proceed solving the problem or to check their solution against a solution implemented by the teacher \citep{udby2016}. The intention of the problems is to provide students with a tool for reflection and self-regulation. 

\section{Methodology for analyzing server logs}
\label{sec:methods}
This section provides an overview of the methodology we have developed and employed to answer the three research questions. The methodology relies on the idea of a \emph{session}: a collection of events, which from the server logs can be tied together by a unique identification code (the session-id). A session consists of a number of time-ordered actions, and has duration, $t_{dur}$. With this definition, we can describe the proposed methodology. The following is a list of all the steps in the methodology. 
\begin{enumerate}
\item Use server logs to create a type of action networks that represent sessions -- we call these session networks. 
\item Find components of behavioral structure
	\begin{enumerate}
	\item Choose structural (network) measures and calculate these for each session network
    \item Perform rotated principal component analysis (PCA) on network measures
	\end{enumerate}
\item Construct network of similar session networks
	\begin{enumerate}
    \item Calculate similarity between each pair of sessions based on distances in rotated PC space
    \item Remove non-significant similarity scores to arrive at a backbone similarity network	
	\end{enumerate}
\item Find clusters of session networks using of community detection in networks
\item Interpret components of structural behavior as well as clusters of similar networks
    \end{enumerate}
    
Next, we briefly describe each step. Details can be found in~\ref{app:detailsMethodology}. 
\subsection{Using server logs to create session networks}
\label{sec:logsToNetworks}
In order to create session networks, we extracted information about sessions and created a table for each as shown in Table~\ref{tab:serverlog}. In that table, type, document\_id, and target\_id uniquely determines the action. Type represents the type of action, for example, whether a hide or show button was pressed, whereas document\_id and target\_id refer to individual pages and specific buttons/links on a particular page. 

\begin{figure}
\includegraphics[width=\textwidth]{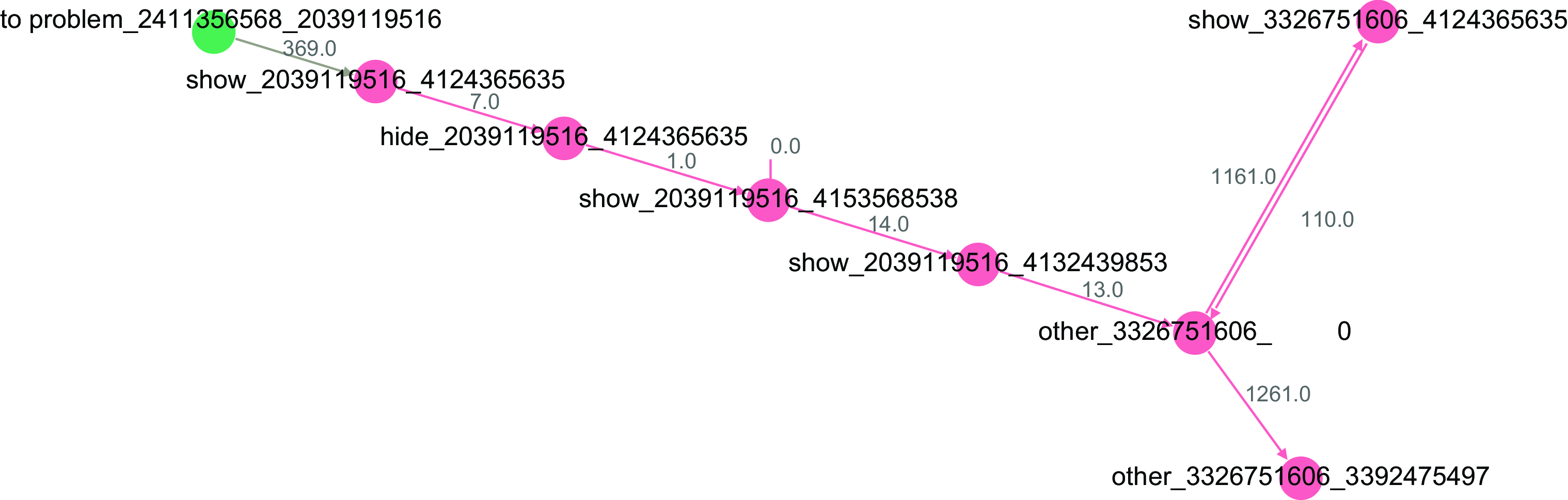}
\label{fig:examplesession}
\caption{Network created on the basis of the events displayed in Table~\ref{tab:serverlog}}
\end{figure}

\begin{table}[h]
\centering
\caption{The information retained for session 1826 after cleaning main server log dump}
\label{tab:serverlog}
\begin{tabular}{lllllll}
 Action & tag & type       & document\_id & target\_id & time stamp & $\Delta t$\\
      \hline
1 & A   & to problem & 2411356568   & 2039119516 & 1411028922 & ...\\
2 & A   & show       & 2039119516   & 4124365635 & 1411029291 & 369\\
3 & A   & hide       & 2039119516   & 4124365635 & 1411029298 & 7\\
4 & A   & show       & 2039119516   & 4153568538 & 1411029299 & 1\\
5 & A   & show       & 2039119516   & 4132439853 & 1411029313 & 14\\
6 & DD  & other      & 3326751606   & 0          & 1411029326 & 13\\
7 & A   & show       & 3326751606   & 4124365635 & 1411030487 & 1161\\
6 & DIV & other      & 3326751606   & 0          & 1411030597 & 110\\
8 & IMG & other      & 3326751606   & 3392475497 & 1411031858 & 1261\\
\hline
\end{tabular}
\end{table}
We now formed networks where nodes represented unique combinations of type, document\_id, and target\_id, and links represented the order. The value of a link was set to the time between actions, $\Delta t$. The resulting session network can be seen in Figure~\ref{fig:examplesession}.

In this way, we created session networks from server logs from three years (2012-2014) involving three iterations of a blended graduate level course on Neutron Scattering. During this period, we have identified a total of 2184 sessions with duration, $t_{dur}>300s$. These session networks represented ways in which students used the wiki textbook. Problem session networks -- session networks in which included a visit to a problem page -- comprised subset ($n=231$) of the total set. 

\subsection{Finding components of behavioral structure}
Inspired by \citet{faust2006comparing}, we aimed at comparing session networks by various network measures. A multitude of network measures exist, and we selected 23 measures of global network characteristics. Some are basic network measures and derivatives, such as the number of nodes, $N$, the number of links, $L$, and the density, $\rho$. Others, such as diameter, $d$, and Target Entropy, $TE$, provide information about the overall structure of the network. Finally, we included connected triads -- or motifs -- in the analysis. Connected triads have been described as the building blocks of networks \citep{milo2002network,milo2004superfamilies}, and may provide more detailed understanding of the structure of session networks. Tables~\ref{tab:measures} and ~\ref{tab:motif} list the 23 measures we included in this study. 

\begin{table}
\caption{Ten of the network measures, which we have calculated in this study. }
\label{tab:measures}
\begin{tabular}{ l p{9cm}}
\hline
  Symbol & Description  \\
  \hline
  $N$ & The number of nodes in the session network. Represent unique actions. \\
  $L$ & The number of (directed) links in the session network. Represents the local timing of actions. \\
    $\rho$ & Fraction of number of links over number of possible links.  \citep{wasserman1994social} \\
      $N_{\leftrightarrow}$ & Number of mutual links.   \\
    $d$ & The diameter of the network. Formally, the longest geodesic (a geodesic is the fewest number of steps needed to connect two nodes) in a network. \citep{wasserman1994social}\\
  $\mean{l}$ & The average path length (average length of geodesics) \citep{wasserman1994social} \\
  $TE$ & Target Entropy, measures the unpredictability of traffic surrounding a node. \citep{rosvall2005networks,bruun2013talking} \\
  $SI$ & Search information, measures the average number of questions needed to navigate from one node to another when following links.\citep{rosvall2005networks,bruun2013talking} \\
  $C$ & Transitivity, the number of closed triangles relative to the number of connected triplets. \citep{wasserman1994social}\\
  $S$ & The entropy of the distribution of number of connections (called the degree distribution). \citep{costa2007characterization}\\
  \hline
\end{tabular}

\end{table}

\begin{table}
\caption{Network motifs we search for. We have used the naming scheme of \cite{milo2002network}. }
\begin{tabular}{ l p{14cm}}
\hline
  Symbol & Motif name and description \\
  \hline
  \includegraphics[scale=0.05]{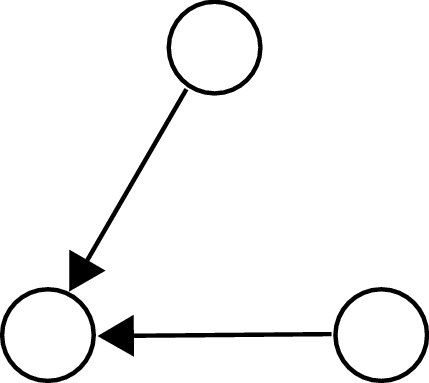} & \emph{V-in}.  Two unique actions, which both precede one unique action at some point during the session.  \\
  \includegraphics[scale=0.05]{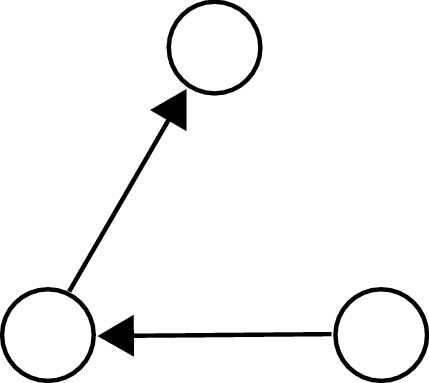} & \emph{Chain}.  Unique actions following each other at some point during the session\\
  \includegraphics[scale=0.05]{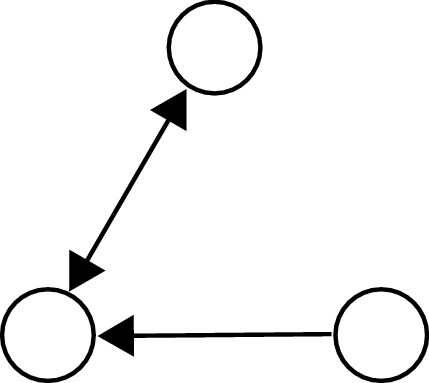} & \emph{Mutual-in}. Two unique actions which followed each other at some point in the session, and one action that preceded one of the unique actions at some point.\\
  \includegraphics[scale=0.05]{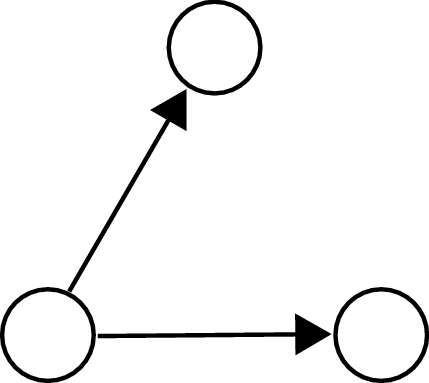} & \emph{V-out}.  One unique action preceding two unique actions at some point during the session.\\
  \includegraphics[scale=0.05]{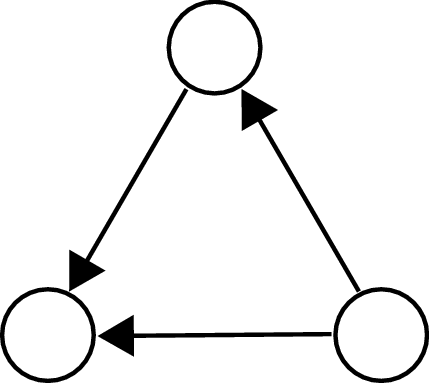} & \emph{Feed-forward-loop}. Like a \emph{chain}, but additionally one action preceding the other at some point.\\
  \includegraphics[scale=0.05]{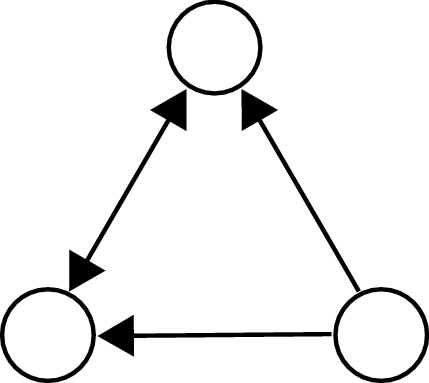} & \emph{Regulated-mutual}. Like \emph{V-out} but with a mutual connection between the two actions. \\ 
  \includegraphics[scale=0.05]{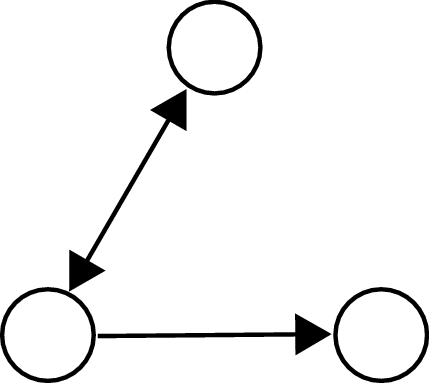} & \emph{Mutual-out}. Like \emph{mutual-in} but with the single link reversed.\\  
  \includegraphics[scale=0.05]{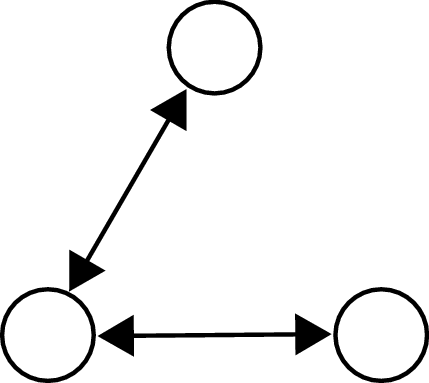} & \emph{Mutual-V}. A combination of \emph{mutual-in} and \emph{mutual-out}. \\
  \includegraphics[scale=0.05]{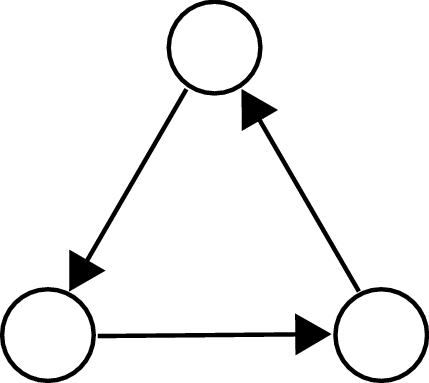} & \emph{Three-loop}. A closed loop of unique actions that followed each other at some point during the session.  \\
  \includegraphics[scale=0.05]{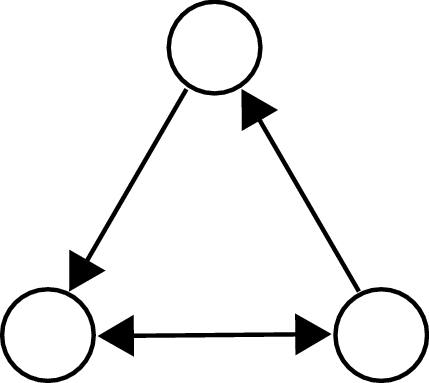} & \emph{Regulated-3-loop}. Like \emph{three-loop}, but with an extra link between two unique actions. \\
  \includegraphics[scale=0.05]{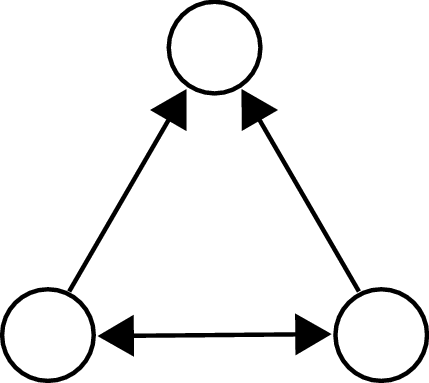} & \emph{Regulating mutual}. Like \emph{regulated-mutual} but with non-mutual links reversed. \\
  \includegraphics[scale=0.05]{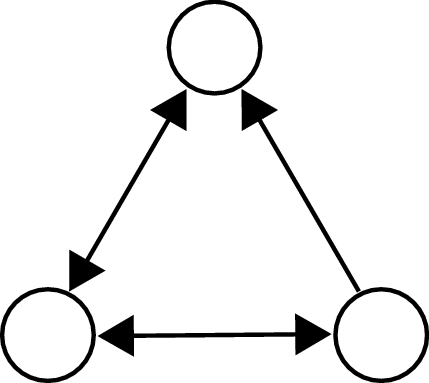} & \emph{Semi-clique}. Like \emph{V-in} with an extra link between two unique events. \\
  \includegraphics[scale=0.05]{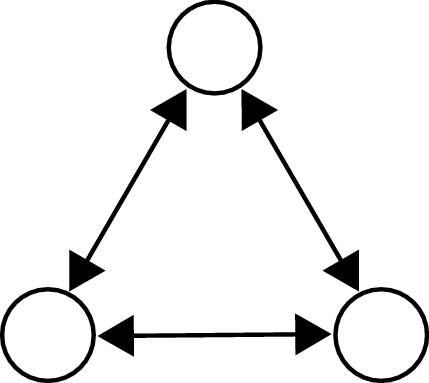} & \emph{Clique}. Three actions that all followed each other at some point during the session. \\  
  \hline
\end{tabular}

\label{tab:motif}
\end{table}
To answer the first research question we performed a PCA on the 23 measures over the 2184 session networks. Principal component analysis (PCA) is a technique commonly used to reduce the dimensionality of a set of variables \citep{james2013introduction,jolliffe2002principal} and have in this respect been used to find similarities in student answers to a questionnaire about their beliefs about online learning \citep{valtonen2009finnish}. The present study used the psych package \citep{revelle2017psych} in the R statistical computing environment \citep{R2017} to find rotated principal components (PCs). We opted to use rotated components since this will "drive loadings towards zero or towards their maximum possible absolute value" \citet[p. 271]{jolliffe2002principal}, thus maximizing differences between components. Thus, we expected each PC to highlight different structural aspects of session networks, and we identify PCs as components of behavioral structure. 

\subsection{Constructing network of similar session networks}
Following \citet{valtonen2009finnish}, we used the Principal Component scores for each network as a basis for similarity. Treating components as spanning a vector space, we calculated the Euclidean distance between each pair of session networks. This procedure produced a symmetric distance matrix. We then converted the distance matrix, $D$, to a similarity matrix, $W$, using the transformation $W_{ij}=\exp{(-D_{ij})}$. Thus, the similarity is a number between 0 and 1, with 1 representing perfect similarity. 

The matrix $W$ is a way to describe a network. In this network, nodes represent session networks and links how similar they are based on distance. Since all session networks will be at a finite distance from each other, the similarity network will be fully connected. Also, a session will be similar to itself. These two conditions make it hard for community detection algorithms to find cluster structure. The standard way to overcome self-similarity is to remove the diagonal. To overcome the finite-distance problem, we follow \citet{brewe2016using} and use local adaptive networks sparsification (LANS) \citep{foti2011nonparametric} to remove insignificant connections. The principle behind LANS is to find out which connections are important for each node. For each link of a node, LANS compares its weight with all other weights of links attached to the node. If the weight is greater than or equals to a predefined fraction of other links the link is kept. Otherwise it is discarded for that node. However, a link can survive if it is significant to just one of the two nodes it connects. To conserve as much information as possible about similarity connections, we choose the predefined fraction so that the resulting network is connected. In doing this, the sparsified similarity network will not consist of isolated islands.   

\subsection{Finding clusters of session networks by use of community detection}
We used the \emph{fuzzy Infomap} algorithm \citep{esquivel2011compression} to partition the similarity network into overlapping clusters. Fuzzy Infomap relies on an information theoretical correspondence between compression and regularity detection. The algorithm can be described as a random walker traversing the network via links. In the similarity network, the walker is expected to spend a lot of time walking between similar session networks, because they are tightly linked. Fuzzy Infomap will exploit the fact that it will be easier to compress information about the walk if similar session networks are grouped into clusters to partition the network. In some cases, a session network will lie on the border between two clusters, and assigning the session network to two or more clusters (with a given percentage belonging to each cluster) will allow for more compression of the information about the walk. The end product of this procedure is an assignment of each session network to one or more clusters.    

\subsection{Interpreting components of behavioral structure}
Each step of the methodology allow for interpretations and characterizations of components of behavioral structures and clusters of sessions. Having characterized components and clusters, the analysis extends in two directions. 

First, we coupled each cluster to non-network measures by searching for over-representation of certain session attributes in clusters, for example, time-of-day, weekday, year, and duration. This was done using the Segregation Z-score, which was also employed by \citet{bruun2014time}. For $Z>1.96$, the Segregation is significantly different from random. 

Second, we chose a representative session network based on raw network measures from each cluster for detailed analysis. Using the session table (see Table~\ref{tab:serverlog}) to trace each student action and time spent between actions this led to a model case for each cluster. The purpose of this was to discern a number of behavioral structures.

\section{Results and interpretations} 
\label{sec:results}
\subsection{Components of behavioral structure}
We performed rotated PCA on the 2184 session networks with $t_{dur}>300s$. The loadings for each rotated component above an absolute threshold value of 0.4 are listed in Table~\ref{tab:pca.loadings}. For further technical details on the rotated PCA, see ~\ref{app:PCA}. Here, we describe each PC as a component of behavioral structure. 

\begin{table}
\caption{Component loadings absolutely above 0.4 for the five components of structural behavior}
\begin{tabular}{ l c c c c c }
\hline
  Network Measure & Comp. 1 & Comp. 2 & Comp. 3 & Comp. 4 & Comp. 5 \\
  \hline
  N & 0.91 & & & & \\
  L & 0.80& & 0.41 & & \\
  $\rho$ & -0.61& & &  & \\
  d & 0.96 & & & &  \\
  $\mean{l}$ & 0.96 & & & & \\
  $N_{\leftrightarrow}$ & & 0.84 &  & & \\
  \includegraphics[scale=0.05]{motif1_V-in.png} & & & 0.81& & \\
  \includegraphics[scale=0.05]{motif2_chain.png} & 0.79& &0.53 & & \\
  \includegraphics[scale=0.05]{motif3_mutual-in.png} & &0.86 & & & \\
  \includegraphics[scale=0.05]{motif4_V-out.png} & & & 0.80& & \\
  \includegraphics[scale=0.05]{motif5_feedFowardLoop.png}  & & & 0.70& & \\
 \includegraphics[scale=0.05]{motif6_regulatedMutual.png}  & & & & & 0.68\\
  \includegraphics[scale=0.05]{motif7_mutualOut.png}  & &0.85 & & & \\
  \includegraphics[scale=0.05]{motif8_mutualV.png} & &0.88 & & & \\
  \includegraphics[scale=0.05]{motif9_3loop.png} & & &0.71 & & \\
  \includegraphics[scale=0.05]{motif10_regulated3Loop.png} & & 0.48& & & 0.48\\
  \includegraphics[scale=0.05]{motif11_regulatingMutual.png} & & & & &0.72 \\
  \includegraphics[scale=0.05]{motif12_semiClique.png} & & 0.66& & & 0.43\\
  \includegraphics[scale=0.05]{motif13_Clique.png} & & 0.52& & & \\
  C & & &  & 0.60& \\
  S & & & &0.82 & \\
  TE & & & &0.83 &  \\
  SI & & 0.54& 0.49& 0.44& \\  
  
  \hline
\end{tabular}

\label{tab:pca.loadings}
\end{table}
\paragraph{Component of behavioral structure 1: Linear Length} 
This component is characterized by a high number of nodes, $N$, and links, $L$, combined with low density and long average path length, $\mean{l}$.
This means that long linear structures in general tend to score high on this component. For example, navigating to a problem with many sub-questions, hints, and model solutions, and pressing all hints and solutions will likely result in high Linear Length. 

\paragraph{Component of behavioral structure 2: Mutuality} 
Mutual links ($N_\leftrightarrow$) load high on this component, as do the motifs which all include mutual links. Consecutively showing-hiding-showing hints and/or solutions will likely contribute to a high Mutuality. 

\paragraph{Component of behavioral structure 3: Navigation}
$V_{in}$, $V_{out}$, chains, and loops load highly on this component. Navigating away from and back to the problem page rather than consecutively repeating a pattern on the same page will likely result in high Navigation. Showing a hint or solution, then searching for information on other pages in the wiki-textbook, and then returning to do some action could be a realization of such a pattern. 

\paragraph{Component of behavioral structure 4: Complexity} 
This is the only component, on which transitivity (a measure of closed motifs \citep{csardi2006igraph}) loads above the 0.4 level. Also, the entropy of the degree distribution and Target Entropy have high loads on this component. This indicates that sessions with high Complexity are probably dominated by many interconnections and shortcuts. Extensive use of the interactive affordances of the wiki-textbook -- navigating to and from hints and solutions, but also to and from pages with information -- is likely associated with high Complexity. 

\paragraph{Component of behavioral structure 5: Erraticism}
The regulating mutual and regulated mutual motifs have large loads on this component and not on other components. Session with high Erraticism likely show a pattern of consecutively performing one action, then another, then first again but also navigating to and from the page where the consecutive actions were performed. The regulating mutual motif would, for example, be seen in cases where a student shows a hint or model solution, hides the hint/solution, tries to solve the problem, shows the hint again, navigates to information in one page, then another page, then goes back to hide the hint again, tries to solve the problem once more and then finally navigates to the first information page to re-check information. 

\subsection{Clusters created of the similarity network}
\begin{figure}
\includegraphics[width=0.8\textwidth]{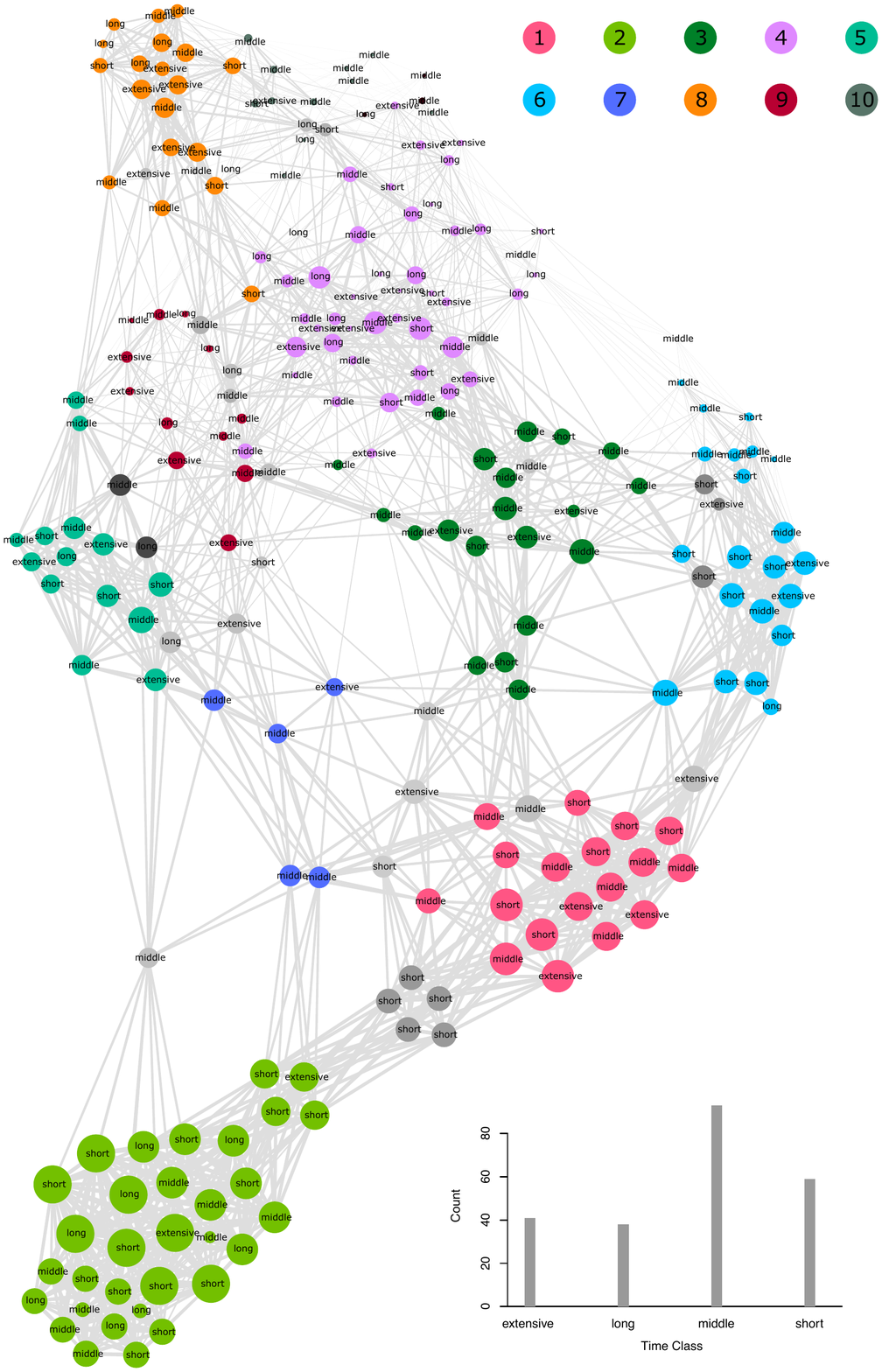}
\caption{The sparsified similarity network. We have found ten overlapping groups. Colors represent clusters; links represent similarity in the space of rotated components. Labels represent time classes found by analyzing length of sessions. The inset shows the distribution of Time Class for all problem solving sessions, see Section~\ref{sec:comparison}. Gray circles belong to more than one cluster.}
\label{fig:simNet}
\end{figure}

Of the 2184 sessions, 231 involved a visit to a problem page. The similarity network was constructed on the basis of these problem session networks. Fuzzy Infomap found 12 clusters in total. However, two of these clusters consisted of one session network each and where discarded from further analysis. Hence, we base the rest of our analysis on the 229 remaining problem session networks. Figure~\ref{fig:simNet} shows the sparsified similarity network with colors highlighting the ten clusters. Sessions in more than one cluster are marked by gray colors. To describe the network, we adopt a map metaphor, so that the upper parts of the similarity network constitute the North and the lower parts the South. Three strong clusters seem to be formed in the South-East "coast" of the network with the clusters 1, 2, and 6 running along the coast. The gray sessions lying between 1 and 2 are shared between these two clusters. North-West from here lays first clusters 3 and 7. They share three sessions, but also share sessions with all other clusters. Further North-Western lay clusters 4,5,8,9, and 10. Cluster 4 is the largest cluster, but also shares many session networks with other clusters. The details of each cluster are shown in Table~\ref{tab:clusterdescription} in~\ref{app:summaryClusters}.  

\subsection{Interpretation of groups of clusters}
\begin{figure}
\includegraphics[width=0.9\textwidth]{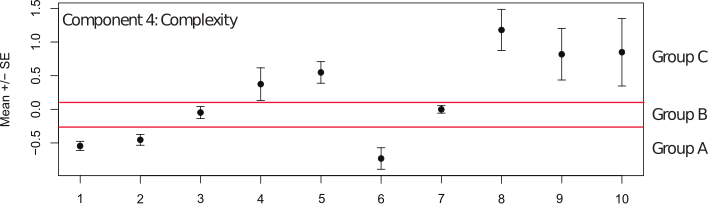}
\caption{We separated clusters into three groups based on their Complexity scores. The red lines indicate separation into Groups.}
\label{fig:complexityComponent}
\end{figure}

For each cluster we calculated the mean scores and standard errors on each component. Based on cluster scores on Complexity, we divided the ten clusters into three groups, each of which represent a different level of complexity. That is, this distinction is based on the mean scores for each cluster on the Complexity Component (see Figure~\ref{fig:complexityComponent}). Below, we provide interpretations of each group.~\ref{app:descriptionGroups} provides further descriptions of each group. 

\subsubsection{Group A -- the least complex group}
The clusters in Group A seem to reflect different instances of the same overall structure: All clusters in this group are very linear, meaning that students only clicks once on each consecutive link. Behavioral structures in this group could reflect that students show hints and solutions and then print the problem and work on Schoenfeld's phases away from the screen; that they use the page as a reference while working; or that they only make use of the read and verify phases of Schoenfeld. 
\subsubsection{Group B -- the medium complex group}
Group B seems to be composed of clusters that exhibit linearity, but with a few 'detours' which we interpret as signs that students are to some extent exploring and interacting with the material and not just activating hints and solutions in succession. Thus, behavioral structures in this group could be a combination of the behaviors of Group A and an Schoenfeld's exploration and perhaps planning phases. In that case detours would involve visiting other pages to gather information. 

\subsubsection{Group C -- the most complex group}
Clusters in this group score highly on Complexity and are quite diverse on the other components (see Figure~\ref{fig:pc_cluster}). This may indicate a number of different strategies. For example, Cluster 10 exhibits high Erraticism, yielding a behavioral structure that could make extensive use of the hint/solution functionality, but maybe not in a strategic way. Such a behavioral structure may signify students continuously carrying out slightly modified plans and verifying solutions and may not evidence self-regulated learning as much as trial-and-error strategies. In contrast, Cluster 4 combines Navigation with Complexity, which may account for deliberate surveying and exploration before carrying out and verifying. This type of behavioral structure might signify the use of a strategy that mimics what an expert would do. Cluster 9 shows high Mutuality which is most likely associated with using the show-hide functionality. The key difference between the Mutuality and Erratic components is the prevalence of regulated/regulating mutual motifs. Thus, the additional clicking associated with the Erraticism is what this analysis structurally associates with an erratic behavior. It is not clear that removing this additional clicking makes behavioral structures with high Mutuality much different from those with high Erraticism. 

\subsection{Comparison with session attributes}
\label{sec:comparison}
In order to further characterize clusters and qualify behavioral structures, we investigated whether time-of-day, day-of-week, week, year, and durations where associated with particular clusters (see ~\ref{app:comparison} for details). Using the Segregation measure \citep{bruun2014time}, we investigated the extent to which attributes were over-represented in clusters. Testing for hour, day, week, and year separately, we found no evidence of Segregation ($Z<1.96$ for all these measures). However, testing for duration of sessions, we did find evidence of Segregation; using the quartiles of a kernel density estimate \citep{R2017,sheather1991reliable} on sessions with $t_{dur}\leq3h$, we were able to categories durations as \emph{short} ($5-25 min$), \emph{middle} ($25min-1.4h$), and \emph{long} ($d1.4h-3h$). Taking three hours as a maximum for one sitting, we labeled sessions with $t_{dur}>3h$ as \emph{extensive}. With this division into time classes, we found significant Segregation on clusters ($Z\approx 4$). We then calculated the per cluster Segregation, finding that clusters in Group A all had a significant overrepresentation of short sessions. Cluster 4 showed an overrepresentation of long sessions. This seems to support the view of this cluster being associated with exploration as well as carrying out and verifying. 

In addition to the Segregation analysis, we also quantified show/hide behavior using the parameter $\mu$, the difference between number of clicks on show and number of clicks on hide relative to the total number of clicks on show and hide. Sessions in Group A featured significantly more show-clicks without subsequent hide-clicks than did Group C. Group B was placed in the middle between the two, but with large uncertainty. ~\ref{app:mu} shows the full analysis.

\subsection{Behavioral structures identified selected sessions}
In answering the third research question, we analyzed a typical session network from each cluster in detail. Using Schonfeld's phases as a guide, the list below summarizes the results of that analysis. Each item represents what we label a behavioral structure. The numbers signify clusters from which the behavioral structures have been inferred. Our full analysis is given in ~\ref{app:sessionExamplesClusters}. 
\begin{enumerate}
 \item \emph{Reading-Selective-Verification}. In problems with more than one sub-problem, students may select particular sub-problems to focus on. This is reflected in a structure where, for example, the first solution is accessed quickly, while solutions to later sub-problems are accessed after spending more time. There is time between checking solutions for off-line work with the problem. 
\item \emph{Reading}. This structure is associated with very little activity. The wiki-textbook is used to look at the formulation of problems and may verify solutions elsewhere. There seems to be little time for online work with the problem. 
\item \emph{Embedded-Reading-Verification}. This structure is like 1. above but embedded in the engagement with another perhaps more complex problem. Here, the solutions to one problem are shown and moved through quickly, while the embedding problem is devoted more time and perhaps complex behavior. 
\item \emph{Read-Verify-Explore}. This structure is characterized by reading the problem after which there is interplay between reading/verifying solutions and exploring the wiki-textbook for information. Time is spent going back and forth between the problem -- where hints and solutions are viewed -- and visiting wiki-textbook pages with relevant information.  
\item \emph{Reading-Peaking}. Here the problem is viewed for some time, after which a solution or hint is shown and then quickly hidden again. Showing the same solution or hint only to quickly hide it again, may happen more than once for each solution and hint.   
\item \emph{Reading-Verification}. This structure is associated with opening all hints and solutions quickly after reading the problem text. 
\item\emph{Embedded-Selective-Verification}. This structure is like as 6. above but embedded in a more complex problem. Here, considerable time is spent before showing particular solutions, while other solutions are accessed more quickly. 
\item \emph{Exploration}. This structure is associated with visiting different wiki-textbook pages, which may not be topically related and subsequently visiting problems, which may not be related to the wiki-textbook pages previously visited. Solutions and hints are likely shown quickly. 
\item \emph{Integrated-Interactive}. This structure is associated with making extensive use of many if not all features of wiki-textbook when engaged with wiki-textbook problems. Textbook pages and problems relate to an initially visited problem are visited, and solutions are shown only after longer periods of time have passed. After showing solutions, some time is spent before the next action. 
\item \emph{Erratic-Interactive} This structure is associated with Erraticism. It involves multiple showing and hiding of hints and solutions with short intervals in between showing and hiding. It may involve going back and forth between related problems, showing and hiding hints and solutions. Thus, like \emph{Integrated-Interactive}, this behavioral structure also makes use of the wiki-textbook affordances but does so in an erratic manner. 
\end{enumerate}

We emphasize that these behavioral structures have been extracted from session networks that showed typical structural characteristics in a cluster. Thus, we do not claim that all or only sessions in, for example, cluster 10 can be characterized as \emph{Erratic-Interactive}. This analysis relied on comparing session network with a session table (like Table~\ref{tab:serverlog}), which held additional information about the time between clicks and the particular pages visited. Thus, behavioral structures hold more information than just network structure. To discern behavioral structure, we also found it necessary to analyze timing between events and to encode knowledge of particular types of actions. 

\section{Discussion} \label{sec:discussion}
We start the discussion by summing up our results and then proceed to discuss the implications of our results for learning behaviors and learning strategies. Then, we discuss the limitations of the study. 

We used rotated PCA of network measures applied to session networks to identify five different components of behavioral structure: Complexity, Linear length, Navigation, Mutuality, and Erraticism. These components of behavioral structures comprise our answer to the first research question. Using the identified components of behavioral structure as a basis for similarity, we constructed a similarity network and used fuzzy community detection to extract ten overlapping clusters of sessions. To answer the second research question, we used cluster mean scores on the five components of behavioral structure to identify three overarching Groups, A, B, and C. Furthermore, we characterized each group in terms of complexity, as well as commonalities and differences within the group. Finally, we used Segregation analysis to show that Group A consisted primarily of short (5-25 min sessions), and furthermore found that Group A also featured more showing of hints/solutions without subsequent hiding than did Group C. In answering the third research question, we have proposed ten behavioral structures by in-depth analysis of a structurally typical session from each cluster. ~\ref{app:overviewFindings} provides a detailed overview of our findings. Here, we turn to discussing behavioral structures and their possible significance to learning behaviors and learning strategies. 

\subsection{Behavioral structures as embedded in learning behaviors and learning strategies}
The networks in this study are embedded in a particular context with a particular meaning, and our focus on only structural aspects hide some of that context. However, there seems to be a dynamic relationship between structure and context; they affect and are affected by each other. On the one hand, using the show-hide functionality has an impact on the structural aspects we can observe. On the other hand, students' intention when engaging with wiki-textbook problems must influence the way they use various functionalities. Our analyses of the sample sessions from Clusters 3 and 7 showed students who seemed to use a solution to a closed problem in their work with another more open problem. As noted above, what could be observed was a linear structure akin to the structures seen in Group A but embedded in a more complex structure. However, as has also been noted, there are many more variables at stake when students engage with learning, which is why the behavioral structures we have identified in this study cannot be labeled learning strategies or even learning behaviors. Rather, behavioral structures form part of what a student does when engaging with the material. One could then argue that if we had access to everything a student does -- online, verbally, and physically -- then a combined map of all these parts would comprise a learning strategy. However, in a constructivist tradition, learning also has to do with \emph{why} one learns and with the context in which one learns. Thus, it is likely that proposed learning strategies, which do not address why students engage with the content or the context of learning, will fail to explain why something was learned and something was not. For example, even if a student 'goes through the right motions' the student may not engage with the material in a meaningful way and may not learn what was intended. We argue instead that a combined map could be seen as a reflection of learning behaviors. 

\begin{figure}
\includegraphics[width=0.6\textwidth]{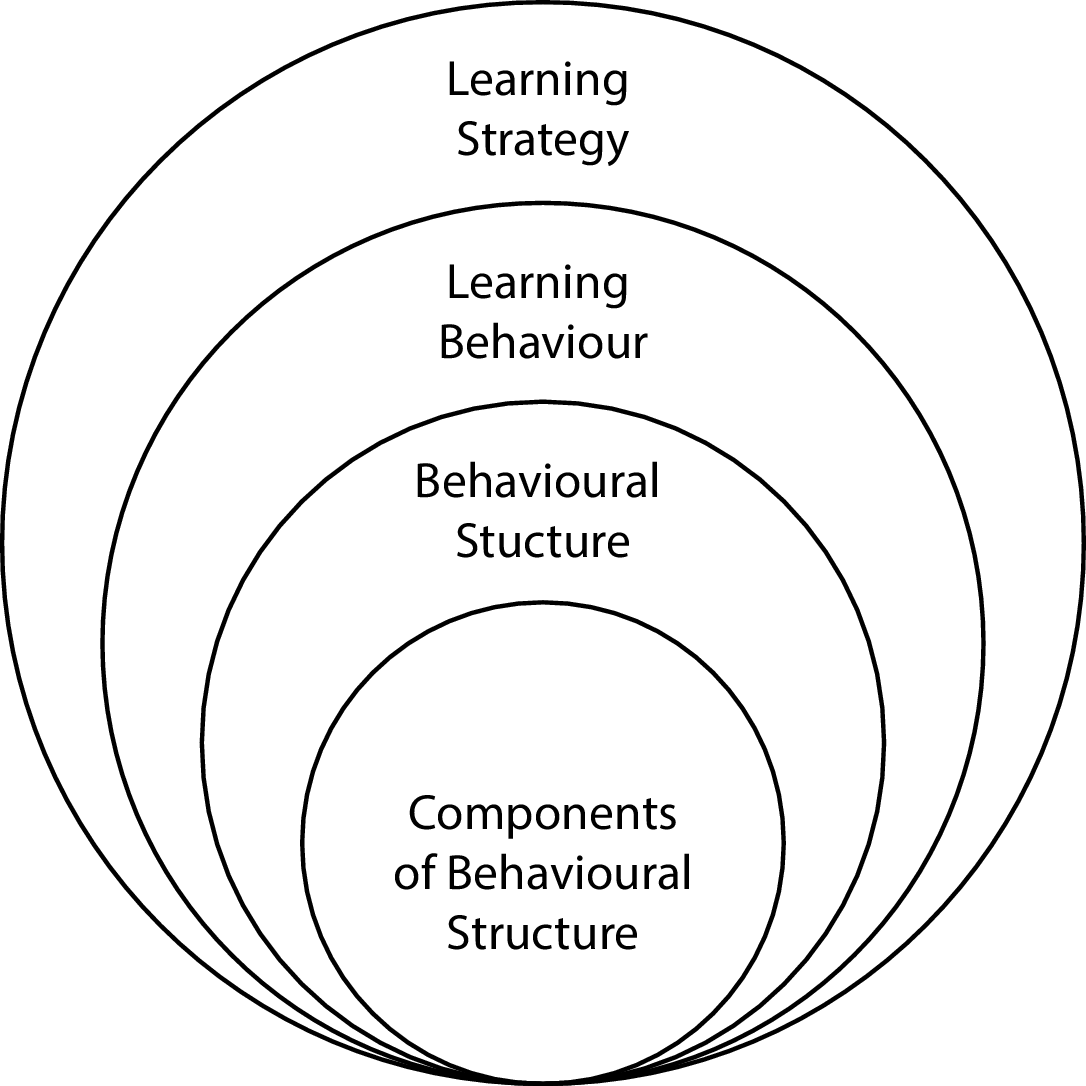}
\label{fig:embeddedStructures}
\caption{Our conception of how components of behavioral structure fit in a larger scheme to make up behavioral structures, learning behaviors, and learning strategies. }
\end{figure}

Figure~\ref{fig:embeddedStructures} sums up this view of learning strategies. Components of structural behavior -- in various modalities -- are combined into behavioral structures. Learning behaviors can then be seen as a combination of many behavioral structures. By adding student intent and context to this interaction picture, we would have a description of learning strategies. Though a detailed discussion of complexity in education \citet{davis2006complexity} beyond the scope of this article, we note that this view is in accordance with the concept of embeddedness.

\subsection{Limitations of the study}
This study has mainly been case study of what can be learned from server logs. There are a number of limitations, and we have identified two types. 

The first type of limitation has to do with the limited modality of our data. The data stems from a web-site, where users were anonymous; we do not know the identity of students. We cannot actually be sure that all problem solving sessions were done by students, even if the activity pertaining to problems was mainly during course weeks (see~\ref{fig:descriptiveM}). This is why we have been careful to describe only behaviors and not students. With non-anonymous data, we would be able to ascribe different sessions to the same student. Thus, we would (a) be able to see if student behavioral structures were stable for a student or if students made use of different behavioral structures at different times and (b) if the same student engaged with two sessions in parallel or just after each other. This might had changed the composition of clusters and groups. Another, but related, limitation is that we cannot know if we are seeing the behavioral structures of only one student. Students may work in small groups when solving these problems. Finally, it is difficult to interpret time between two consecutive actions. Longer times may be due to a student thinking really hard, a student using a different modality (e.g. pen and paper), or a student taking a break. These limitations are the reason we have focused on behavioral structures, but this focus in turn also limits our inferences. 

The second set of limitations has to do with the sample. While 231 problems may seem like a lot, the number of students who engaged with these problems is probably in the order of 40-50. The clusters we have found may not represent a larger student body. Also, the interactions we investigated are specific to the area of Neutron Scattering and may be biased. 

\section{Conclusion}
Starting from 2184 session server logs of student interactions with a wiki-textbook on neutron scattering, we created session networks, which captured the structure of student online actions. Using rotated principal component analysis, we identified five components of behavioral structure: Complexity, Linear Length, Navigation, Mutuality, and Erraticism. Based on session network scores on these components, we created a network of similar sessions, and found three large groups -- A, B, and C -- with different overall characteristics. Group A showed linear behavior and an overrepresentation of short (5-25 min) sessions, while Group C showed more complex behavior involving, for example, more hiding of hints and solutions after showing them. We took Group A to represent interactions which did not make much use of the interactive affordances of the wiki-textbook, while Group C represented interactions which made more use of these affordances. Finally, we analyzed typical sessions in clusters to identify ten behavioral structures, and we have argued that these structures can be seen as integral parts the behaviors that students employ when learning in a blended setting. 

\clearpage
\bibliographystyle{model5-names}
\bibliography{refs,selfAnon}

\clearpage
\section*{Acknowledgements}
We would like to thank the students following the Neutron Scattering Course 2012-2014 for letting us perform research on their user data.
This research was funded by the European Union via the SINE2020 project (GA no. 654000).

\clearpage
Highlights
\begin{itemize}
\item The paper presents a novel approach of analysing server logs of student interactions with a web-site
\item The approach conceptualises student interactions with a web-site as behavioural structures.
\item The paper identifies behavioural structures as these appear in connection with problem solving in an online textbook format.
\item The paper identifies and characterises three over-arching groups of behavioural structures: linear, complex, and in-between. 
\item Linear structures are linked to shorter session durations and less use of web-site affordances than complex structures. 
\end{itemize}
\clearpage

\appendix
\section{Additional Background}
\label{app:additionalBackground}
\subsection{Additional research on hints and solutions in problem solving}
\label{app:addResProbSolv}
However, the provision of hints and solutions will not automatically lead to students learning how to solve problems. \citet{chi1989self} find that students' way of engaging with solutions in an intervention study is correlated with their success in a post-test. They find that "good students" (sic) exhibit behaviors, which are different from "poor students". In their analysis, a "good student" elaborates on the given example solutions while studying them to generate own understandings of the laws of physics. They tend to monitor their own understanding and this self-monitoring generates observable actions. For \citeauthor{chi1989self} actions are student utterances. In the case of monitoring web-based actions, the observable actions might be visiting text pages, other similar problems, or showing and hiding hints and solutions. 

The use of example solutions is not trivial in the sense of either student maturity or gender. Using a self-reported survey, \citet{sandelin2011} find that some pharmaceutical students employed deep learning strategies when studying worked out examples and others did not. They find a student maturity effect on learning strategies; fourth year students were much more likely to use example solutions for deep learning than first or second year students. Furthermore, they find a gender difference. Females were more likely to use examples productively than males. 

\subsection{Additional research on clustering}
\label{app:addResClust}
Clustering methods assume underlying distance metrics and distribution of variables. For example, in analysing an on-line questionnaire about Finnish high school student' readiness of adoption of on-line learning, \citet{valtonen2009finnish} use principal component analysis to identify underlying variables (labelled sub-scales in that study) and subsequently $k$-means clustering \citep{dutt2015clustering} to find three distinct groups of students with different attitudes towards adopting on-line learning. In education as well as in other fields, clusters are often identified using either $k$-means or hierarchical clustering \citep{dutt2015clustering}. $k$-means takes as the input the number of clusters to be identified and subsequently finds clusters based on each observation's distance in some mathematical space to a centroid. Hierarchical clustering (either agglomerative or divisive) is based on the distance  between each pair of observations. Clusters are identified on the basis of some cut-off distance. $k$-means clustering is often seen as a non-costly and quick way of acquiring clusters. The number of clusters needs to be specified beforehand and the output is an unstructured set of clusters. This limits subsequent analyses. Hierarchical clustering adds information about structure, since observations are joined at different levels of dissimilarity. However, once two observations have been joined, they are treated as one, which can result in the loss of information about structure. Furthermore, there is no standard for identifying clusters in hierarchical clustering.

\subsection{Wikis as textbooks}
\label{app:addResWiki}
Digital and on-line textbook have been conceptualized in many different ways. Some digital textbooks are published in a software framework which offers the reader further functionalities such as e.g. highlighting of words, commenting and dictionary look-ups (most eBooks) but also and enhancing visualization and illustration of concepts by figures with zooming option or 3D animations (e.g. iBooks). Other textbook-like formats are HyperPhysics \citep{hyperPhysics} and documents generated automatically from texts \citep[see e.g.][]{latex2html}. 

A novel possibility is to use the wiki-format to create a textbook edited by experts within a particular knowledge domain. This would entail a constantly updated textbook, with the functionality of a wiki. Such a textbook would not entail collaboration between many disparate sources like Wikipedia. Rather, it would be the focused enterprise of scientists with very specific knowledge. By far most of the educational research investigates the use and potential of wikis as student collaborative platforms \citep{augar2004,parker2007,lin2009,matthew2009,karasavvidis2008}, whereas educational research on the use of teacher-produced wikis as teaching material for students seems to be lacking.

A wiki-textbook would share the affordance of all digital textbooks that words and concepts are structured to be easily accessible and searchable. On the other hand, reading longer texts on a computer screen presents students with an increased cognitive load because of the need of scrolling, which in turn makes it more difficult to locate previous information when it is needed. This tends to reduce reading comprehension \citep{waestlund2005,singer2017reading}. Also, while a digital textbook may afford highlighting and note-taking, it lacks the flexibility and robustness of pen and paper \citep{karlsson2017myth,holman2005paper,piper2009tabletop}. For example, pen and paper are not dependent on battery and may be hurled to the floor without suffering fatal damage. Thus, allowing a combination of a digital and printed versions of a textbook may be an optimal solution for accommodating diverse student preferences. 

Software underlying a wiki-textbook interface will log user actions. Large amount of data is stored and that data describes the behaviors of users of the wiki-textbook. 

\section{Additional information the wiki textbook as implemented in this study}
\label{app:addInfoWiki}

In this section we describe the creation and maintenance of the wiki-textbook. We also provide examples of text and problems as they appear in the wiki-textbook. 

The wiki-textbook is based on a textbook-like file (for brevity, we refer to that file as the textbook) featuring content and problems provided by eight experts within the field of neutron scattering. One expert acts as a main author and moderator of chapter coherence, chapter organization, and topical relevance. The textbook is continuously updated when important new findings are found. The tasks of the main author represent specific challenges to a collaborative writing project \citep{lin2008} versus other wiki-related projects. Therefore, the development of the textbook and transformation into the wiki-textbook format are two separate processes. 

\begin{figure}
\label{fig:wikistructure}
\includegraphics[width=\textwidth]{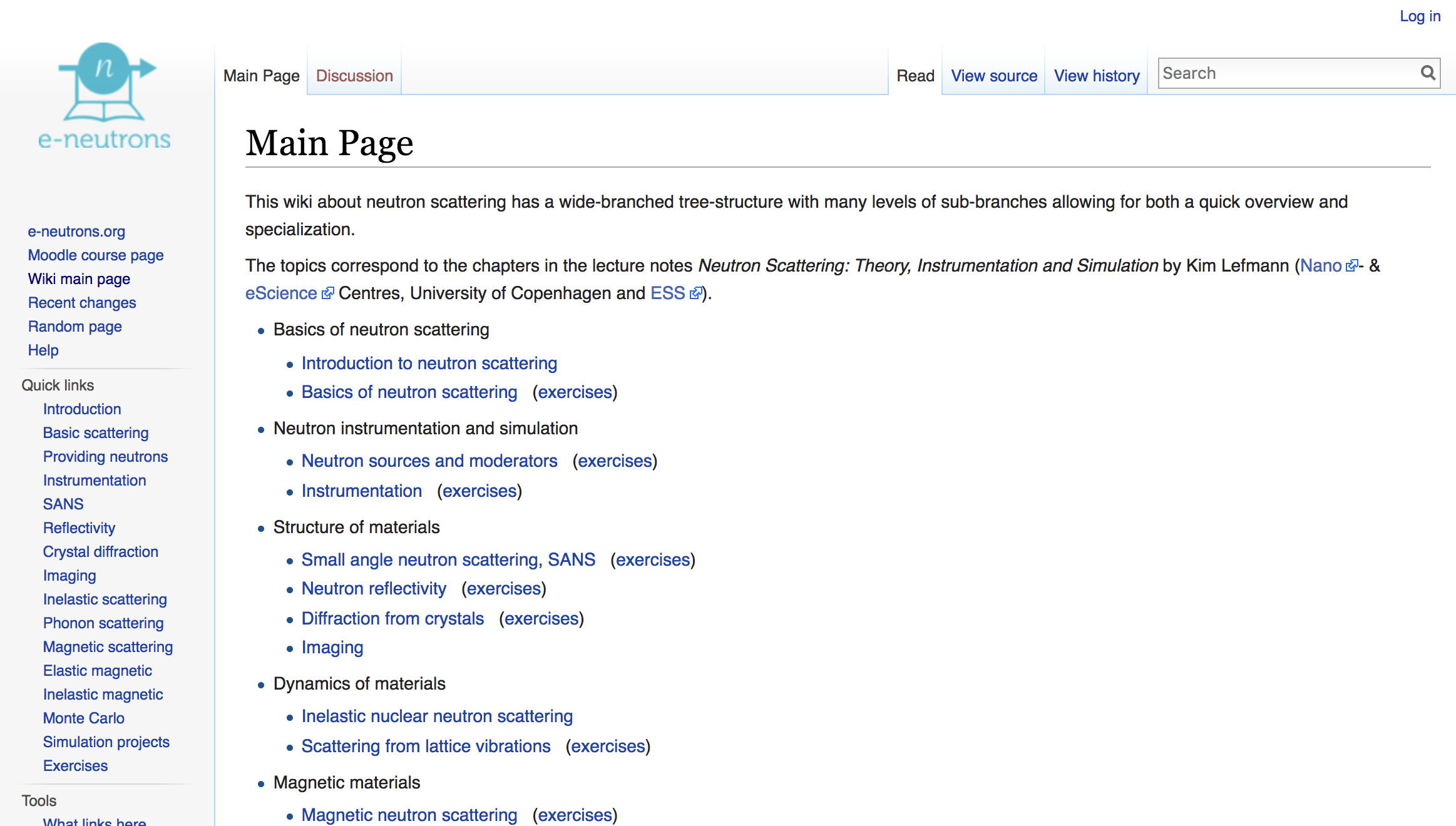}
\includegraphics[width=\textwidth]{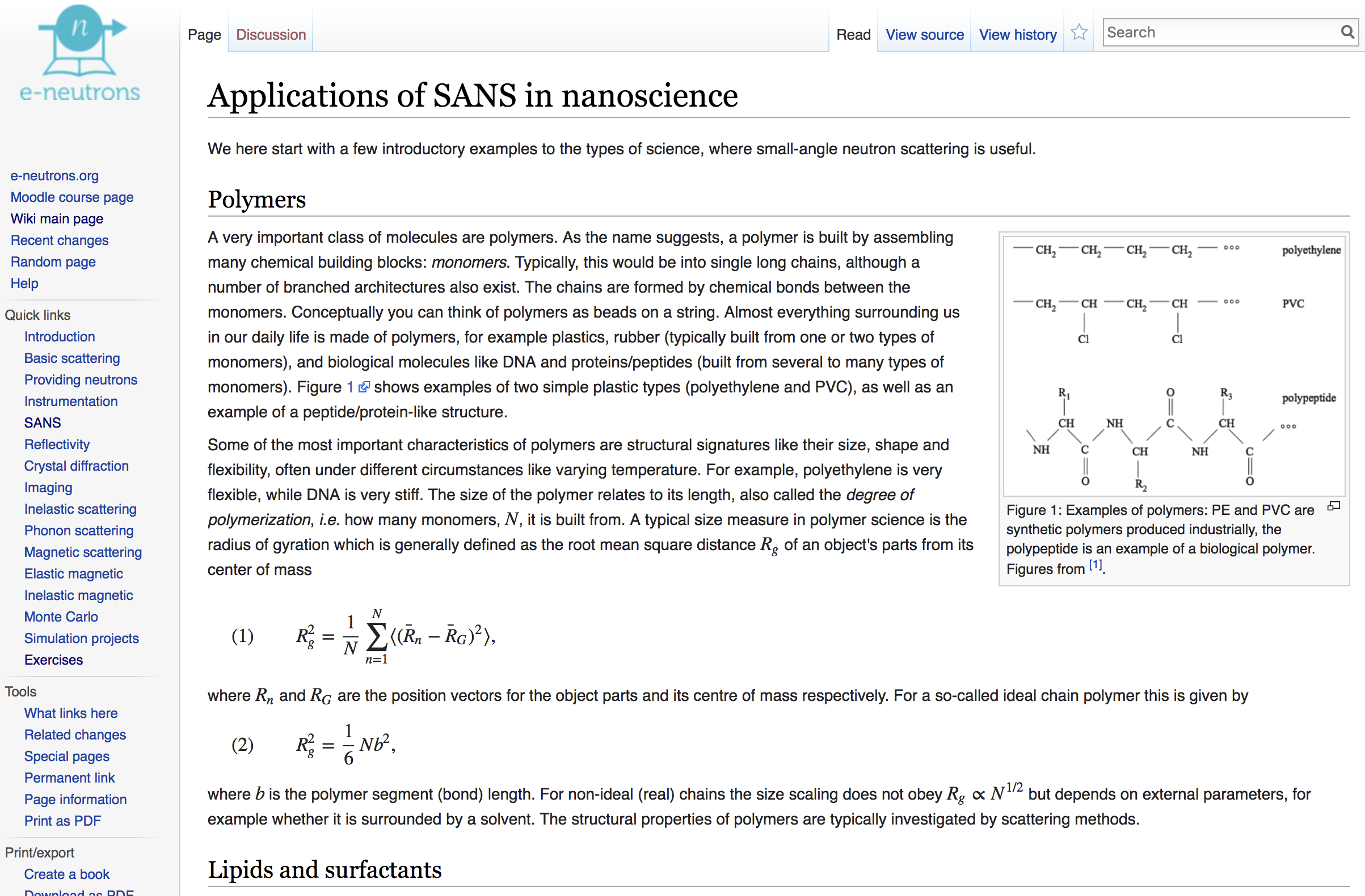}
\caption{A screen shot showing the main page of the wiki-textbook (top) and an example of text in the wiki text book (bottom)}
\end{figure}

As described in the main article, the wiki-textbook is organized in a tree-structure that mirrors the textbook. Figure~\ref{fig:wikistructure} shows an example. 
All text chapters are listed in a menu at the main page of the wiki-textbook and the problems are placed in a separate chapter. See Figure~\ref{fig:wikistructure}. The reason for this structure is to limit the need for scrolling, since text-passages become smaller for each subdivision. 

In the transformation from textbook to wiki-textbook, hints and model solutions are added. The hints and solutions have been developed separately over roughly five years taking typical student procedural and technical (mathematical) questions for solving each problem into account. Consistent with the finding that solving many of the same types of problem need not lead to better problem solving skills \citep{kim2002students}, each problem has to do with a particular concept or situation that is relevant to neutron scattering. Furthermore, the users are students at the graduate level, which suggests that they likely will employ deep learning strategies when using hints and solutions \citep{sandelin2011}. Thus, unlike learning material employed in related research \citep{pol2005solving,harskamp2007schoenfeld,pol2008effect,pol2009}, the wiki-textbook does not focus on development of student problem solving strategies.

The hints and solutions to problems in our study are in practice implemented with several extensions to the MediaWiki engine, specifically ShowHide, MathJax and CrossReference, whereof the last two are related to the contents rather than the functionality. We have used the MediaWiki engine to create the wiki-textbook, and make use of multiple extensions and functionalities. Specifically, ShowHide, MathJax, and CrossReference \footnote{See MediaWiki documentation at \url{https://www.mediawiki.org/wiki/MediaWiki}}. The wiki-textbook is part of a larger on-line package of teaching materials for neutron scattering, which may be accessed at \href{https://www.e-neutrons.org}{e-neutrons.org}. The software as well as access to the web-site is free. 

\section{Details on methodology}
\label{app:detailsMethodology}
\subsection{Definitions: sessions and networks}
\label{app:definitions}
\paragraph{A session} 
A collection of events, which are tied together by a common identification code, called the session-id. For each user action a number of identifiers are recorded (see Section 5.1 in main text). One of those identifiers is a time stamp. This is the number of seconds from some starting time, $t_0$, likely when the server was started. In identifying the duration of sessions, the first time stamp, $t_{start}$, was subtracted from the last time stamp, $t_{end}$, to calculate the duration of the session in seconds, $t_{dur} = t_{end}-t_{start}$. We infer that all unique sessions involve the use of one computer. However, they can extend through a long period of time and can involve more than one student. 

The methodology operates with two levels of networks, the level of the individual session and the level of similarity between individual sessions. Since the methodology is based on network analysis, the next step is to define nodes and links for the two levels. 

\paragraph{Session network level} 
At this level, we define a node as an action, $A_i$, that occurs on a web-page. For example, if a [show] button is clicked on a particular problem page. Each such action is recorded in the server log with a time stamp, $t_{i}$. A link, $L^{k}_{ij}$ is created between two actions, $A_i$ and $A_j$, if $A_j$ follows $A_i$ in the server log, that is if $t_{j}>t_{i}$. The time between actions is then, $\Delta t^{k}_{ij}=t^{k}_{j}-t^{k}_{i}$. A link can occur multiple times, which is represented by the index $k$. 

\paragraph{Similarity network level} 
At this level, each node represents a session network, and links represent similarity. As will be shown later, the methodology uses Euclidean distances in a space spanned by principal components of structural network measures as a basis for similarity. Having created a similarity network, a community detection algorithm \citep{esquivel2011compression} will partition this network into clusters of structurally similar sessions.\\

\subsection{Details on creating session networks}
\label{app:logsToNetworks}
As described in Table 1 in the main text, we used \emph{document\_id} to uniquely identify a viewable document, for example, a page containing a problem and \emph{target\_id} to identify the target of the action. Targets can be hyper-links to other pages, show/hide buttons, or other actions (such as showing an image, scrolling, or dragging). \emph{Tags} are HTML-tags. To distinguish between different types of actions, we created an identifier called \emph{type}. Using URL names, the Document Object Model \citep{DOM} of the wiki-software, and HTML-tags, we could use the type-identifier to label whether an action was navigation to a wiki-page with a problem or not, if it was a click on show or on hide, or other. Pages that involve wiki-textbook problems all have the word \emph{problem} in the URL. We used this to create a list 231 sessions, which included events on pages with one or several of 25 wiki-textbook problems. We subsequently used this list to keep track of these problem solving sessions. Finally, we used the timestamps for each session to calculate each time period, $\Delta t^{k}_{ij}$, between two subsequent actions, $A_i$ and $A_j$.

Each action, $A_i$ was now labeled with a code consisting of $type_i$,$document\_id_i$, and $target\_id_i$. For example, in Table 1 in the main text, the first and second events are
$$A_{1}=to\quad problem\_2411356568\_2039119516$$
$$A_{2}=show\_2039119516\_4124365635$$
with the link
$$L^{1}_{1,2}=369s$$
These two actions are now represented as nodes in the network in Figure 2 in the main text, as are the rest of the actions and time differences. Thus, each session is visualized by nodes and links between nodes. The length of the arrows and the size of the nodes have no significance in this representation.

\subsection{Short descriptions of network measures}
\label{app:networkMeasures}
In this section, we give a brief description of the network measures used in this study. We have already described nodes, $N$ and links, $L$. The density, $\rho$, can be seen as a measure of how many unique events are connected; in a high density session network a user have navigated between a large fraction of the unique events. Mutual links,  $N_{\leftrightarrow}$ indicate that a student has navigated back and forth between two unique actions. 

Shortest paths from unique actions to other actions are central to the next four measures. A shortest path between two nodes, $A_i$ and $A_j$ is the least number of possible other nodes one has to visit in order to get from $A_i$ to $A_j$ \citet{wasserman1994social}. Shortest paths are often used in navigational models. The average (shortest) path length, $\mean{l}$, is a measure of how long paths one will usually have to take. The diameter, $d$, is the longest shortest path in the session network and is often taken as a linear measure of the size of the network. Target entropy, $TE$ and search information $SI$ \citep{rosvall2005networks,bruun2013talking} use shortest paths to gauge the activity around an action and how difficult it is to find $A_j$ from $A_i$, respectively. In this study, we calculate $TE$ and $SI$ for the whole network. Doing this for $TE$ will yield information about the predictability of a session network. For example, if a session network consists of a linear string, $TE=0$, while it will be higher for more complex networks. $SI$ will yield information about the navigability of the network; a low SI will signify easy navigation from action to action on average. 

Motifs \citet{milo2002network,milo2004superfamilies}, which are the thirteen possible different connected triads of nodes have been described as the building blocks of networks and in our case they may reflect patterns of action. Sessions with a relatively large number of \emph{chains} would be very linear, while session networks with a high number of \emph{cliques} (six directed links between three nodes) would signify a lot of navigating back and forth between pages and/or showing/hiding  hints/solutions. We expect to find more of certain motifs in session networks depicting particular kinds of behavioral structures. 

\subsection{The LANS procedure}
\label{app:LANS}
In order to make the similarity network more amenable to cluster analysis, we follow \citet{brewe2016using} and use local adaptive networks sparsification (LANS) \citep{foti2011nonparametric} to remove insignificant connections. The principle behind LANS is to find out which connections are important for each node. For a link $L_{ij}$, LANS compares its weight, $W_{ij}$ , with all other weights of links attached to the node. If $W_{ij}$ is greater than or equal to a predefined fraction of other links,  $f=1-\alpha$, where $\alpha$ can be interpreted as a p-value, the link is kept. Otherwise it is discarded for that node. However, a link can survive if it is significant to just one of the two nodes it connects. As a rule of thumb, we choose the smallest $\alpha$-level where the network is still connected. In doing this, the sparsified similarity network will not consist of isolated islands. In contrast to the directed and possibly weighted session networks, the similarity network is undirected and weighted.

\section{Additional information on results}
\label{app:addInfoRes}
\subsection{Technical details on PCA}
\label{app:PCA}
We performed the rotated principal component analysis on the 2184 session networks with $t_{dur}>300s$. Of these, 231 was identified as problem solving sessions. The parallel analysis suggested that five principal components would be sufficient for our purposes. Running the PCA with five components and varimax rotation we found that each component accounted for 8\% or more of the variance (adding up to a total of 77\%). Adding and extra rotated component accounted for an additional 5\%, and we chose to keep five components. The loadings for each rotated component above an absolute threshold value of 0.4 are listed in Table 4 in the main article. We identified each rotated component as a component of behavioral structure, and session networks with high scores on a component should exhibit network measures in accordance with the loadings on that component. 

We correlated session scores on each component and found no correlations ($p>0.9$ for all correlations). However, for the subset of 231 problem solving sessions we did find small but significant correlations between scores on some of the components. For the 231 problem sessions, component 1 and 2 were negatively correlated with Component 4 ($r=-0.2, p<0.01$ and $r=-0.3, p<10^{-6}$, respectively). Component 2 and 4 were positively correlated with component 5 ($r=0.2, p<0.01$ for both). Thus, the problem sessions seem to represent a special kind of behavior. In following paragraphs, we describe each component and give preliminary characterizations of the underlying behavior. We limit our characterization to behavior associated with problem solving, although the components have been extracted from a broader set of sessions. 

\subsection{Summary of information about clusters}
\begin{landscape}
\begin{table}
\caption{Description of the 10 overlapping clusters. The first row shows number of sessions, with [..] indicating number of sessions with full membership in the cluster. Apart from the first row, the numbers are weighted averages and uncertainties on last digits.}
\begin{adjustbox}{width=1.4\textwidth}{!}
\begin{tabular}{ l c c c c c c c c c c }
\hline
  Name & Clus 1 & Clus 2 & Clus 3 & Clus 4 & Clus 5 & Clus 6 & Clus 7 & Clus 8 & Clus 9 & Clus 10 \\
  \hline
  $N_{ses}$ & 28 [18] & 36 [30]  & 30 [18] & 61 [42] & 19 [13] & 28 [21] & 13 [5] & 25 [18] & 23 [12] &20 [10]\\
  N & 9.0(7) & 4.4(6) & 13(2) & 19(3)& 4.7(7)& 19(4)& 5(3)& 6(1)& 9(2)& 12(6)\\
  L & 8.3(7) & 3.6(5) & 14(3)& 25(5)& 5.0(8)& 19(4)& 5(3)&7(1) & 12(3)& 20(11)\\ 
  S & 0.49(3) & 0.55(7) & 0.6(1)& 0.83(8)& 0.9(1)& 0.32(6)& 0.4(2)& 0.7(2)& 0.8(2)& 0.8(4)\\ 
  $\rho$ & 0.100(6) & 0.19(3)& 0.054(9)& 0.067(7)& 0.23(5)& 0.045(9)& 0.08(3)& 0.16(5)& 0.10(5)& 0.06(3)\\ 
  d & 8.1(7) & 3.6(6) & 10(2)& 11(2)& 3.6(5)& 17(4)& 4(2)& 5(1)& 6(2)& 8(5)\\ 
  $\mean{l}$ & 3.3(3) & 1.8(2) & 4.0(7)& 4.2(5)& 1.7(2)& 6.5(1.4)& 1.6(9)& 2.0(4)& 2.3(6)& 2.6(1.3)\\ 
  $\leftrightarrow$ & 0.4(3) & 0.03(6) & 1.8(7)& 5(2)& 2.2(4)& 0.4(4) & 1.4(8)& 1.0(6)& 6(2)& 6(4)\\ 
  TE & 0.010(8) & 0 & 0.067(15)& 0.16(2)& 0.18(4)& 0.015(9)& 0.04(2)& 0.16(5)& 0.20(7)& 0.20(10)\\ 
  SI & 0.04(5) & 0.00(1) & 0.7(2)& 2.0(3)& 0.3(1)& 0.2(1)& 0.2(2)& 0.5(2)& 1.2(4)& 1.7(9)\\ 
  C & 0(0) & 0(0) & 0.01(1)& 0.13(2)& 0.003(5)& 0.006(7)& 0(0)& 0.29(7)& 0.07(4)& 0.15(7)\\ 
 \includegraphics[scale=0.05]{motif1_V-in.png} & 0.04(7) & 0 & 0.8(3) & 5(2)& 0.1(1)& 0.5(4) & 0.02(4) & 0.9(3) & 0.9(7)& 4(2)\\ 
 \includegraphics[scale=0.05]{motif2_chain.png} & 7.1(7) & 2.7(5) & 12(3)& 23(5)& 1.9(7)& 18(4)& 3(2) & 4(1)& 5(2) & 15(7)\\ 
 \includegraphics[scale=0.05]{motif3_mutual-in.png} & 0.1(1) & 0(0) & 1.0(4) & 7(3) & 1.1(2) & 0.3(2) & 0.7(4) & .8(5) & 4(2) & 6(4)\\ 
  \includegraphics[scale=0.05]{motif4_V-out.png}  & 0.04(7) & 0(0) & 0.7(3) & 4(1) & 0.01(2) & 0.3(3) & 0(0) & 0.5(2) & 0.6(6) & 3(2)\\
  \includegraphics[scale=0.05]{motif5_feedFowardLoop.png}  & 0(0) & 0(0) & 0.02(4) & 0.6(3) & 0(0) & 0.03(6) & 0(0) & 0.1(1) & 0.01(3) & 0.3(3)\\ 
  \includegraphics[scale=0.05]{motif6_regulatedMutual.png}  & 0(0) & 0(0) & 0.08(8) & 0(0) & 0(0)& 0(0)& 0(0)& 0(0) & 0.3(5) & 0.6(4)\\ 
  \includegraphics[scale=0.05]{motif7_mutualOut.png}  & 0.1(1) & 0.01(3) & 0.7(3) & 6(3) & 0.6(0.3) & 0.2(0.2)& 0.5(0.5) & 0.6(0.4) & 3(2) & 5(3)\\ 
  \includegraphics[scale=0.05]{motif8_mutualV.png}  & 0.02(3) & 0(0) & 0.06(8) & 3(2) & 0.2(2) & 0(0) & 0.2(2) & 0.1(1) & 2(1) & 2(2)\\ 
  \includegraphics[scale=0.05]{motif9_3loop.png}  & 0(0) & 0(0) & 0.05(8) & 1.2(3) & 0(0) & 0.1(1)& 0(0) &0.8(2) & 0.1(2) & 0.9(6) \\ 
  \includegraphics[scale=0.05]{motif10_regulated3Loop.png}  & 0(0) & 0(0) & 0(0) & 0.6(4) & 0.01(2) & 0(0)& 0(0) & 0.2(2) & 0.2(2) & 0.7(6)\\
  \includegraphics[scale=0.05]{motif11_regulatingMutual.png}  & 0(0) & 0(0) & 0(0) & 0(0) & 0(0) & 0.02(3)& 0(0) & 0(0) & 0(0) & 0.5(4)\\
  \includegraphics[scale=0.05]{motif12_semiClique.png}  & 0(0) & 0(0) & 0(0) & 0.2(2) & 0(0) & 0(0)& 0(0) & 0(0) & 0.3(2) & 0.5(7)\\ 
  \includegraphics[scale=0.05]{motif13_Clique.png}  & 0(0) & 0(0) & 0(0) & 0.04(5) & 0(0) & 0(0)& 0(0) & 0(0) & 0.2(2) & 0.2(2)\\ 
\hline
\end{tabular}
\end{adjustbox}

\label{tab:clusterdescription}
\end{table}
\end{landscape}

\begin{figure}
\includegraphics[width={0.75\textwidth}]{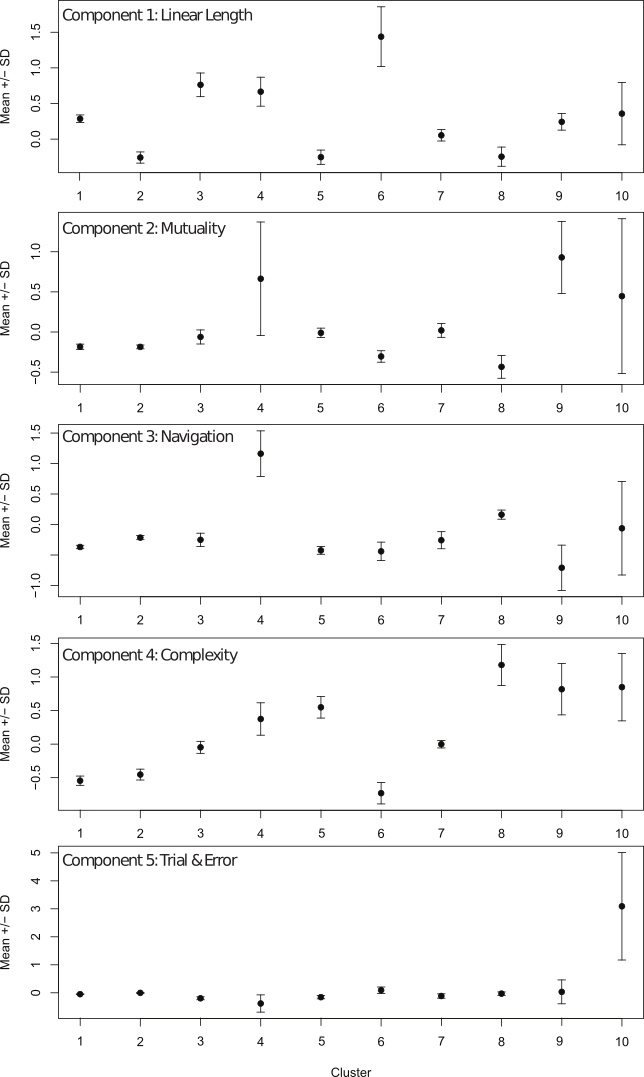}
\caption{\label{fig:pc_cluster}Mean component scores per cluster. Error bars represent 95\% confidence intervals.}
\end{figure}
\subsection{Descriptions of Groups}
\label{app:descriptionGroups}
\subsubsection{Group A -- the least complex group}
\emph{Description}: A total of 82 sessions are represented in this group. 75 of these sessions are full members of Group A, meaning that they are either fully part of clusters 1, 2, or 6 or their membership is shared between clusters 1, 2, and 6.\\
The clusters in Group A have significantly negative scores on the Complexity Component. Thus, Group A could be said to represent the least complex, or most linear, networks. Within this group,  Linear Length separates the clusters: Cluster 6 scores highest of all clusters on Linear Length, Cluster 1 scores medium, and Cluster 2 scores low on Linear Length, so the linear length and average path length $\mean{l}$ varies within this group where session networks in Cluster 6 contain more nodes and links and session in Cluster 2 fewest. This can be confirmed by consulting Table ~\ref{tab:clusterdescription}.

\subsubsection{Group B -- the medium complex group}
\emph{Description}: There is a total of 40 sessions in Group B, 23 of which are full members.\\
Group B score approximately 0 on the Complexity Component, which is significantly different than group A and C, i.e. sessions of group B has medium complexity. \\
Group B is also specifically separable on Target Entropy - higher than group A but lower than Group C. See Table~\ref{tab:clusterdescription}.
Clusters 3 and 7 within group B is are separated by Linear Length. Cluster 3 scores significantly higher, meaning the linearity length parameter (as well as the average path length $\mean{l}$) is bigger in Cluster 3. Table~\ref{tab:clusterdescription} confirms that Cluster 3 scores higher on every network measure that loads on Linear Length, while they score similarly on TE, S, and C.  

\subsubsection{Group C -- the most complex group}
\emph{Description}: There is a total of 123 sessions represented in this group, 107 of which are full members.\\
Mean scores on the Complexity Component in Group C are significantly positive. Group C thus represents the most complex sessions of the dataset. Within group C, the clusters are further characterized on their mean scores on other components as described below.
\begin{itemize}
\item Cluster 9 scores significantly positive on the Mutuality Component (but comparable to Clusters 4 and 10, which have large error-bars). I.e. these session networks show higher Mutuality than other sessions in Group C.
\item Cluster 4 scores significantly higher than other clusters on the Navigation Component. Table~\ref{tab:clusterdescription} that sessions in Cluster 4 have more $V_{in}$, $V_{out}$, and loop motifs when compared to other clusters. This accounts for the higher scores on Navigation.
\item Cluster 10 scores significantly positive on Erraticism but has large within group variation. Consulting Table~\ref{tab:clusterdescription}, this seems to be due to the large number of \emph{regulating/regulated mutual} motifs.
\item Cluster 5 and 8 both score low on Linear Length. In that sense they are similar to Cluster 2 in Group A, but with a much higher score on Complexity. Clusters 5 and 8 score significantly different on the Mutuality Component (Cluster 5  approximately 0, Cluster 8 negative) and the Navigation Component (Cluster 5 negative, Cluster 8 slightly positive). 
\end{itemize}

\subsection{Descriptive statistics}
\label{app:descriptiveStat}
\begin{figure}[!h]
\includegraphics[width=\textwidth]{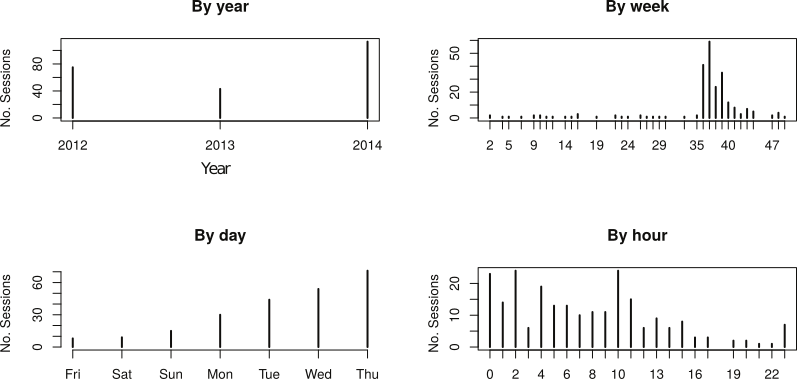}
\label{fig:descriptiveM}
\caption{Descriptive statistics for various measures related to time.}
\end{figure}

Descriptive statistics for various measures related to time are shown in Figure~\ref{fig:descriptiveM}

\subsection{Distributions of duration times per cluster}
\label{app:duration}
\begin{figure}[h!]
\includegraphics[width=0.8\textwidth]{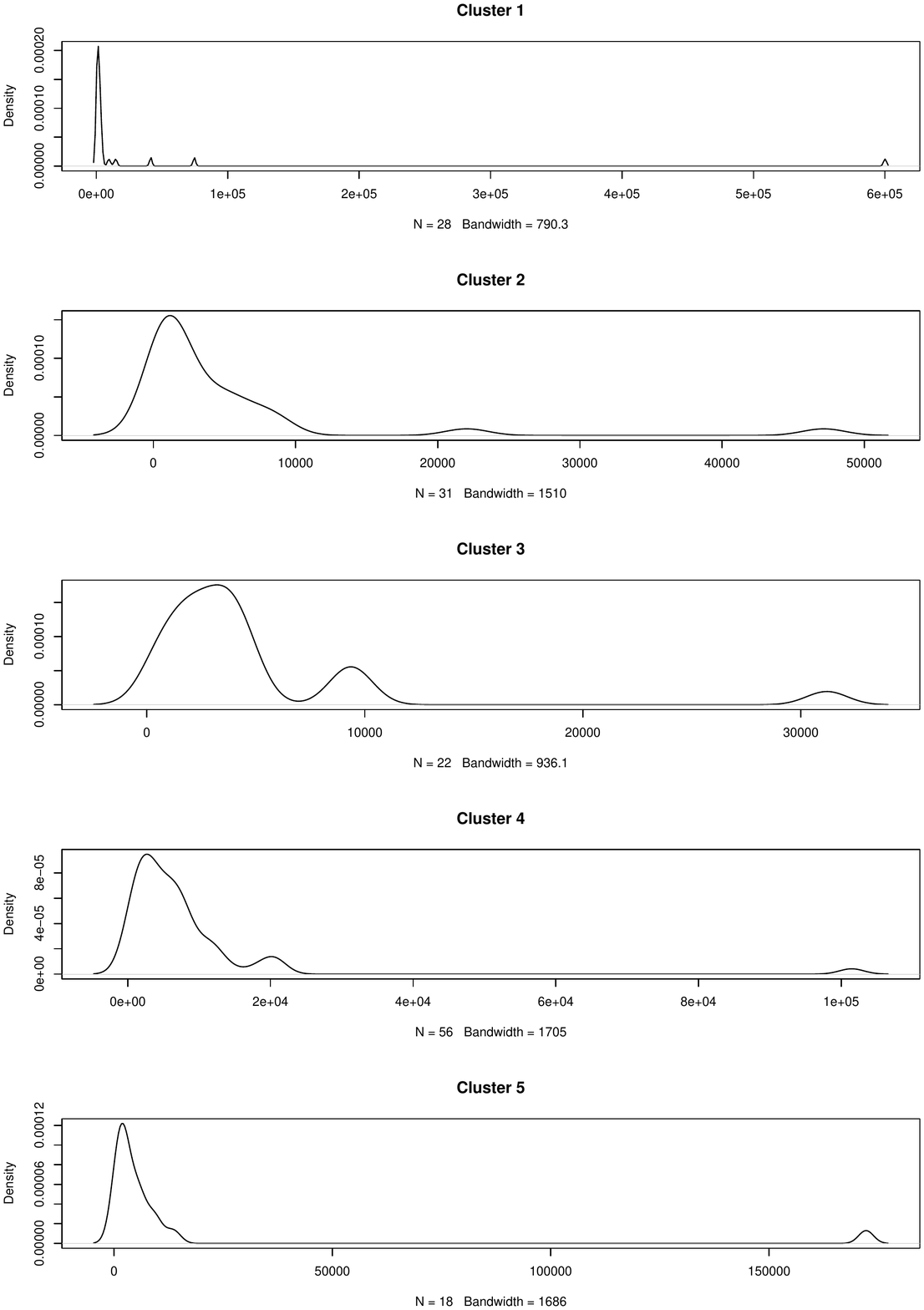}
\label{fig:durations1-5}
\caption{Distribution of session durations in Clusters 1-5.}
\end{figure}
\begin{figure}[h!]
\includegraphics[width=0.8\textwidth]{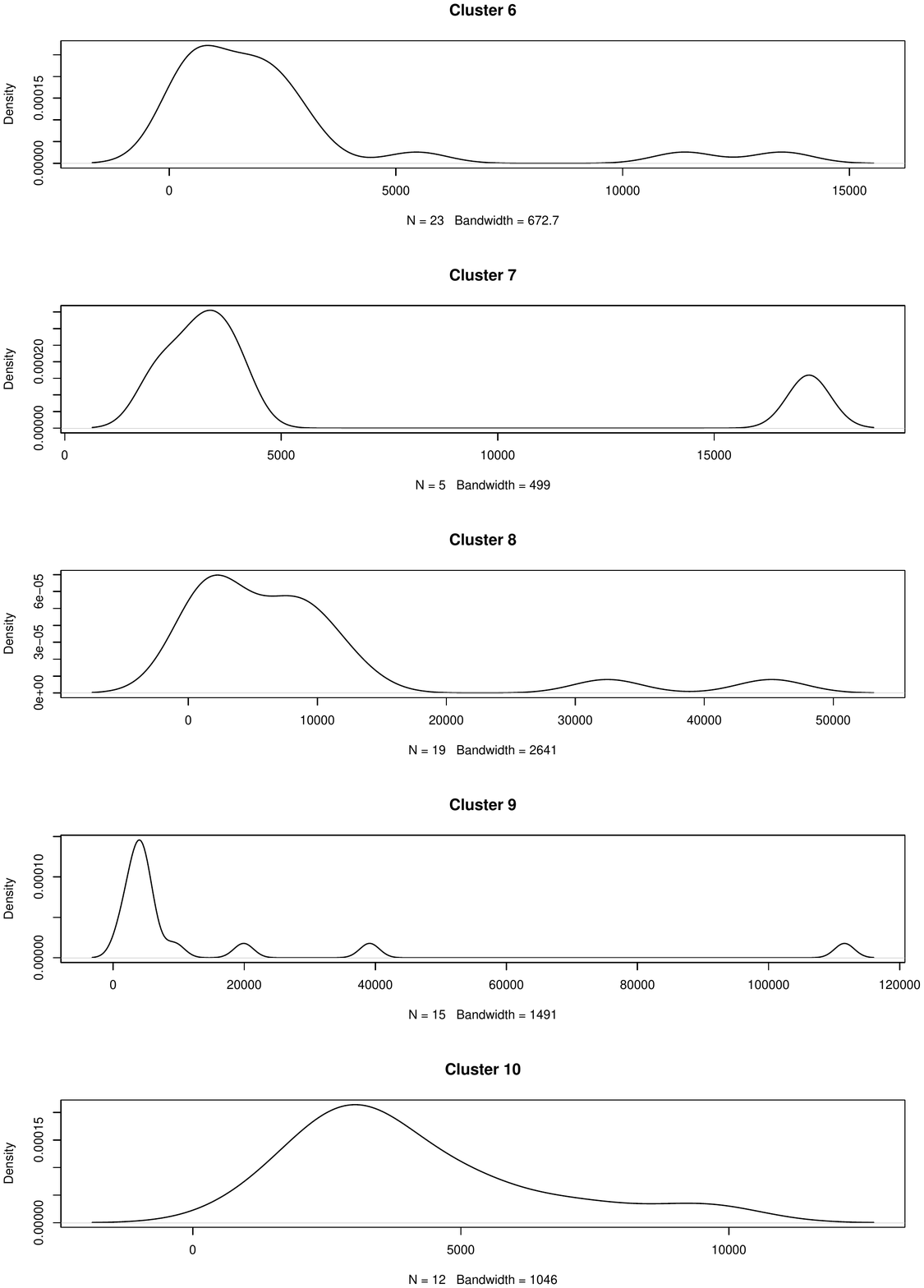}
\label{fig:durations6-10}
\caption{Distribution of session durations in Clusters 6-10.}
\end{figure}

\subsection{The Segregation measure}
\label{app:segregation}
We used the Segregation measure as employed by \citet{bruun2014time}. The Segregation measures any over representation of a particular node attribute in a group. For example, a cluster might consist of sessions that happened more at a particular week day than could be randomly expected. This would result in a high Segregation when compared to random variation. We follow \citet{bruun2014time} and assign attributes randomly and calculate the Segregation for the random assignment. We do this $N=10^4$ times and calculate the Z-score $Z=(X-<X>)/\sigma_X$. For $Z>1.96$, the Segregation is significantly different from random \cite{bruun2014time}. The Segregation has been developed for hard partitioning only, while we employ a fuzzy partitioning. To accommodate this deficit, we created a hard partitioning, in which a session network was put in the cluster to which it had the highest percentage of belonging. For example, Session Network 50, is 80\% part of Cluster 4 and 20\% part of Cluster 10. In the hard partition, Session Network 50 would then be assigned to Cluster 4. 

\subsection{Comparison with session attributes}
\label{app:comparison}
\begin{figure}
\includegraphics[width=\textwidth]{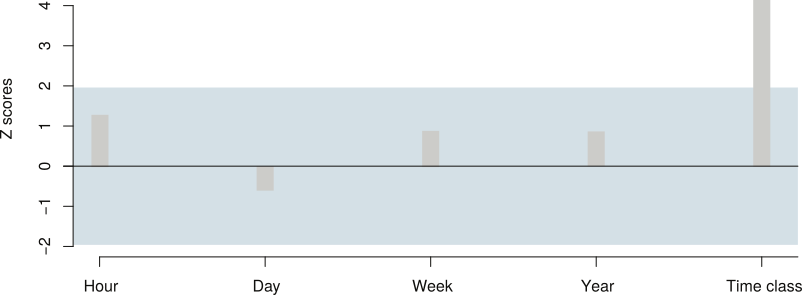}
\label{fig:z-scores}
\caption{Segregation Z-scores based on $N=10^4$ random assignments. $Z>1.96$ (above shaded region) indicates significant Segregation. We only find significant segregation for the time class variable. }
\end{figure}

In order to further characterize clusters and qualify behavioral structures, we investigated whether time of day, day of week, week, year, and durations where associated with particular clusters (for descriptive statistics of these variables, see Appendix ~\ref{app:descriptiveStat} above. Using the Segregation measure \citep{bruun2014time}, we investigated the extent to which attributes were over-represented in clusters. Testing for hour, day, week, and year separately, we found no evidence of segregation ($Z<1.96$ for all these measures, see Figure~\ref{fig:z-scores}). 

Testing to see, if particular durations of sessions were associated with particular clusters posed some difficulty because in general, the distributions of durations were not comparable (see Appendix~\ref{app:duration}). Instead, we created a Time Class variable by dividing durations into a discrete set of time classes: \emph{short}, \emph{middle}, \emph{long}, and \emph{extensive}. We did this by first dividing the durations into two smaller sets, one with session durations less than 3 hours (201 sessions) and the other set with session durations of more than 3 hours. This division was made based on our expectation that 3 hours would for most be the maximum time spent in one sitting and then the remaining time was probably spent on breaks from the actual work in the wiki-textbook (without logging out). Sessions with durations of more than 3 hours were labeled \emph{extensive}. For the remaining sessions, we made a kernel density estimate \citep{R2017,sheather1991reliable} and used the quartiles as separators. From this, the duration of \emph{short} sessions is from 5 to about 25 minutes, \emph{middle} sessions range from about 25 minutes to 1.4 hours, while \emph{long} sessions range from 1.4 hours to 3 hours. With this division, we found segregation in the data set on the time class variable. See Figure~\ref{fig:z-scores}.

To further characterize clusters, we calculated the Segregation per cluster (see Table~\ref{tab:segregation}, finding that clusters in Group A (Clusters 1, 2, and 6) all showed an over-representation of short sessions. 

Cluster 3 showed an over-representation of middle duration sessions. This may indicate that students exhibiting this behavioral structure use some of their time to explore information on the wiki-textbook, but primarily use the problem page as a reference for the text and for checking hints and solutions. 

Cluster 4 showed an over-representation of long duration clusters. This seems to support the view that in this cluster we would find sessions with deliberate surveying and exploration before carrying out and verifying. The added characterization then is that  when engaging with problems in an area that is likely novel, students that employ this behavior use substantial time to do so.

\begin{table}
\label{tab:segregation}
\caption{Per cluster distributions of duration categories in clusters with significant Segregation on duration categories.}
\centering

\begin{tabular}{lccccc}
\hline
Cluster & Short & Middle & Long & Extensive & Sum  \\
&5-25 min.& 25 min.-1.4h&1.4-3h&3h+ &\\
\hline
1       & 0.46  & 0.36   & 0 & 0.18 & 1      \\
2       & 0.45  & 0.26   & 0.23 & 0.06 & 1       \\
3       & 0.18  & 0.64   & 0    & 0.18 &  1      \\
4       & 0.14  & 0.34   & 0.30 & 0.21 &   1     \\
6       & 0.48  & 0.39   & 0.04 & 0.09 &    1    \\
\hline
All   & 0.26  & 0.40   & 0.16 & 0.18   & 1    \\ 
\hline
\end{tabular}

\label{tab:segregationTime}
\end{table}

\subsection{Patterns of showing and hiding hints and solutions}
\label{app:mu}
\begin{figure}[h!]
\includegraphics[width=0.5\textwidth]{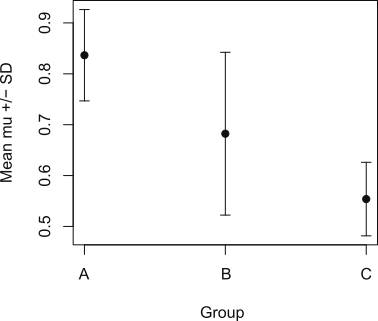}
\caption{\label{fig:mu_group} $\mu$ scores versus group. The higher the $\mu$ score the lower the tendency to click on hide.}
\end{figure}

We speculated that the over representation of short sessions in Group A might be attributed to a particular kind of on-line behavior -- one of showing hints and solutions. Conversely, we speculated that the complex behaviors in Group C would be associated with more use of showing and hiding.  Investigating this assumption, we calculated $\mu$, which gauges the relative amount of show/hide clicks for each session
\begin{equation}\label{eq:mu}
\mu_s=\frac{N_{show}-N_{hide}}{N_{show}+N_{hide}}
\end{equation}
A value of 0 would mean the same amount of clicks and hides and a value of 1 would mean no hides. Some sessions involved no clicks on show, and they have been excluded in the analysis below, leaving 186 sessions. We found that Group A had a significantly higher $\mu$ than Group C (\ref{fig:mu_group}. This is interesting because we now know that behavioral structures in Group A are associated with (1) showing hints and solutions without re-hiding them and (2) that they take 5-25 minutes, which is ample time to read the hints and solutions to many problems. 
An average value of $\mu$ for each cluster $\mean{\mu}$ is found by weighing the value of each session $\mu_s$ with the partial membership of each cluster ($m_s$=1 if session s is fully in the cluster and m=0 if it is not in the cluster at all) and then taking the average of the weighed $\mu$s for each cluster,
\begin{equation}\label{eq:muav}
\sigma_{mu}=\sqrt{\frac{\sum^{s\in C} m_s\mu_s}{\sum^{s \in C}m_s}}
\end{equation}
The standard deviation and confidence intervals were calculated on the weighted means:
\begin{equation}\label{eq:musd}
\mean{\mu}=\frac{\sum^{s\in C} m_s(\mu_s-<\mu>}{\sum^{s \in C}m_s}
\end{equation}
Finally, the uncertainties were calculated as
\begin{equation}\label{eq:muunc}
\Delta=\frac{\sigma}{\sqrt{\sum^{s \in C}m_s}}
\end{equation}
Confidence intervals in figures is $CI=1.96*\Delta$.

\subsection{Session examples in clusters }
\label{app:sessionExamplesClusters}
\subsubsection{Example of sessions from group A (Clusters 2, 1 and 6)}
\label{app:examplesA}
We have analyzed each of the example session networks below with the help of a table akin to Table 
\subsection{Example of sessions from group A (Clusters 2,1, and 6)}

\begin{figure}[h!]
\includegraphics[width=0.5\textwidth]{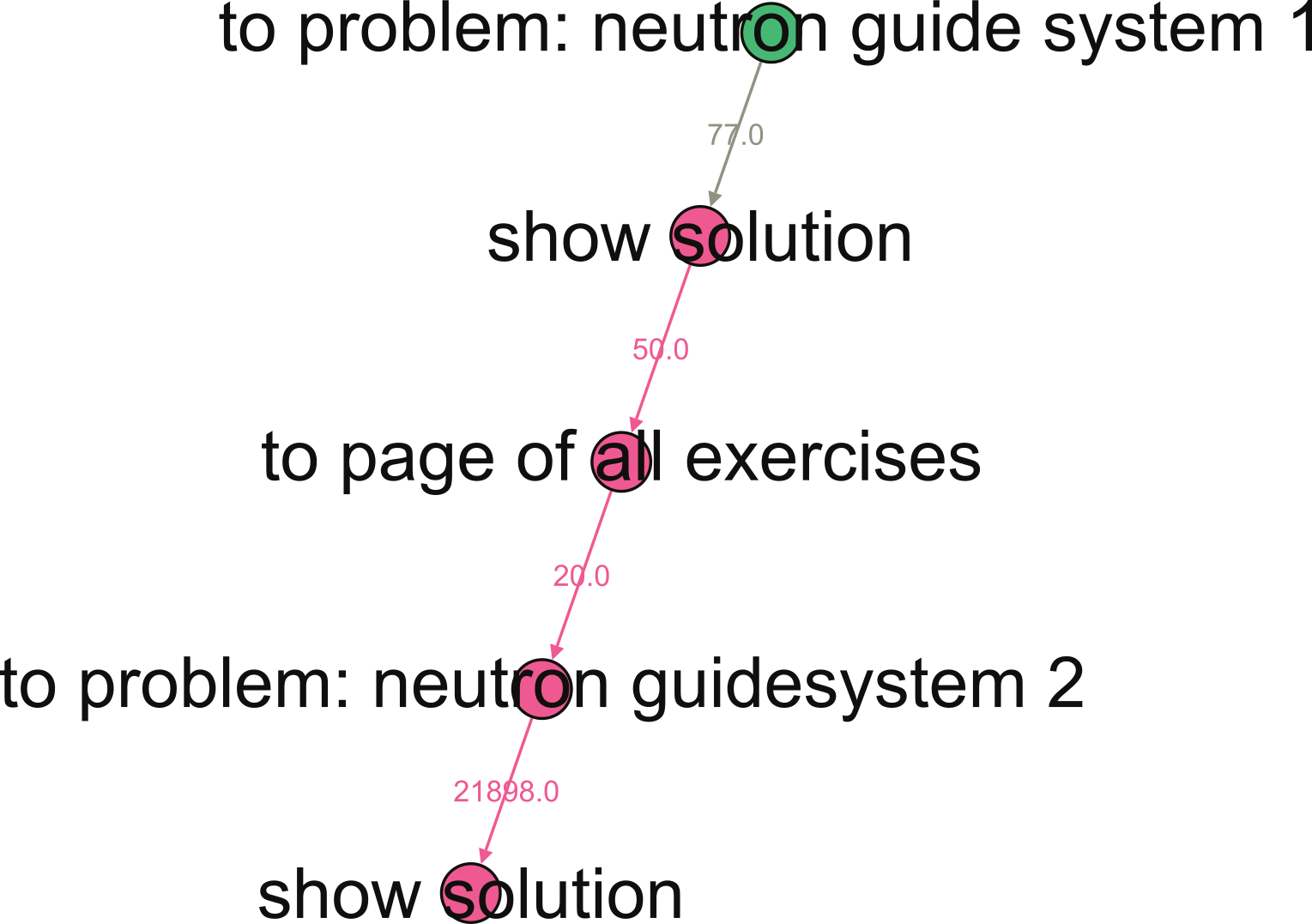}
\caption{Session 1675 showing a typical example of a session network in Cluster 2.}
\label{fig:clu02_1675}
\end{figure}

\begin{figure}[h!]
\includegraphics[width=0.5\textwidth]{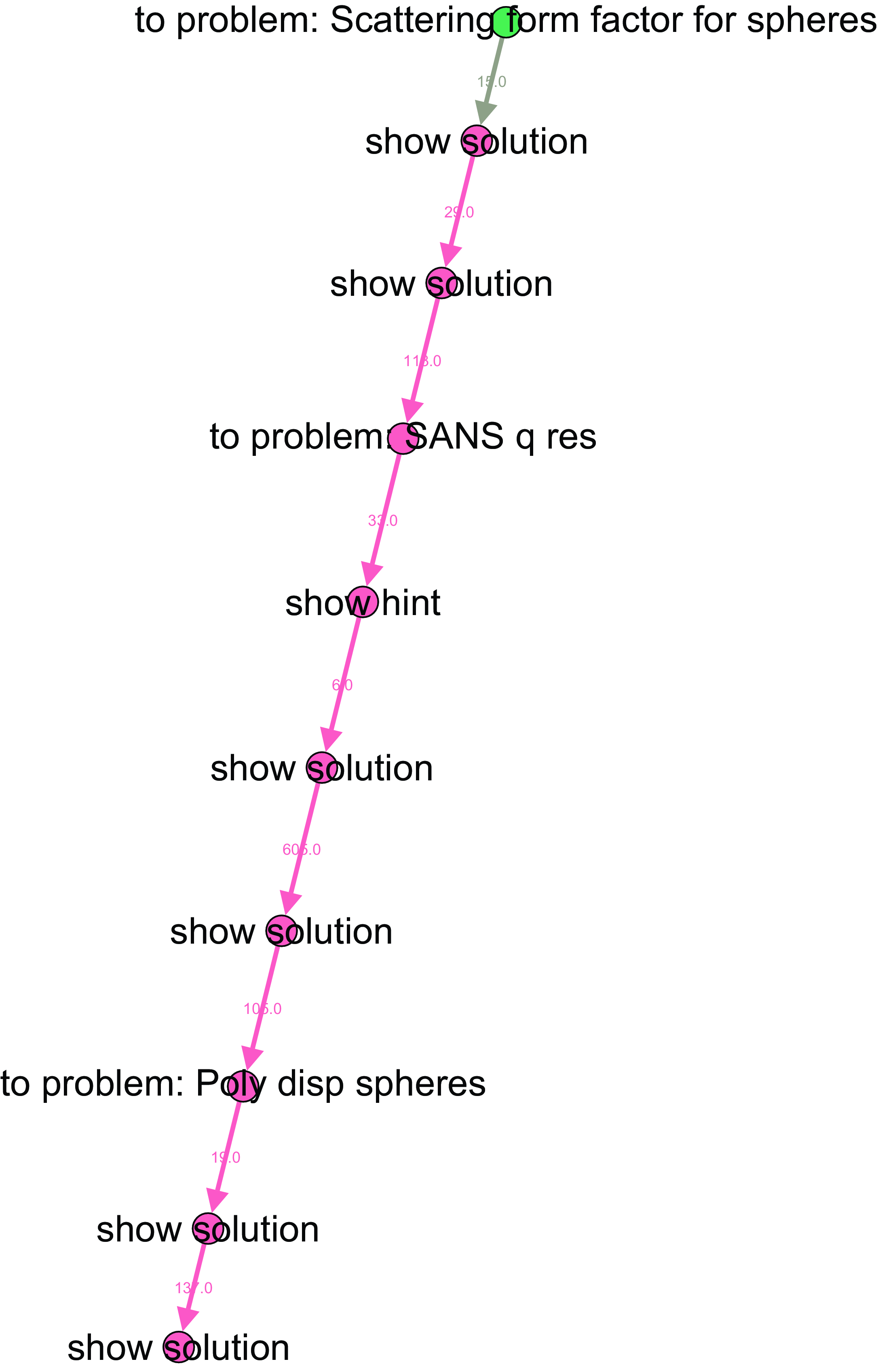}
\caption{Session 1046 showing a typical example of a session network in Cluster 1.}
\label{fig:clu01_1046}
\end{figure}

\paragraph{Cluster 2}Figure \ref{fig:clu02_1675} shows a typical example of a session network from Cluster 2 with low Linear Length in the least complex group (A). \\
\textit{Description}: The session is 6 hours and 7 minutes long but only has 5 nodes. The student starts at an exercise page and after $\sim 1$ minute he opens the solution to one of the questions. He spends close to minute presumably reading the solution and then goes to the overview page of all exercises where he after 20 seconds clicks on another exercise which is topically closely related to the first. After roughly 6 hours where he presumably works with the second exercise (the particular exercise involves simulation and can thus be lengthy to solve) he shows a solution to a question in this exercise.\\
\textit{Interpretation}: The session shows very little activity in the wiki-textbook probably due to activity elsewhere. The wiki-textbook is mostly used to look at formulation of exercises and check solutions after considerably working with them elsewhere.\\
The strategy of this learner seems to mainly use the wiki-textbook to read exercise formulations and verify solutions worked out elsewhere. There is very little use of interactive features of the wiki-textbook and no looking for info in text pages. This behavioral structure could be called \emph{Read}.
\paragraph{Cluster 1} Figure \ref{fig:clu01_1046} shows a typical example of a session network in Cluster 1 which has medium Linear Length within the least complex group (A). \\
\textit{Description:} The session has 10 nodes and takes 17 minutes.
In this session the student starts at an exercise, looks at it for 15 sec, skips the hint and instead clicks on the solutions to the two questions of the exercise which he shows 29 sec apart. He then then goes to another exercise, opens the hint to the first question after half a minute and solution shortly after, but then spends 10 minutes (presumably working on the second question) before showing the solution to the second question. After a couple of minutes he then goes to a third exercise, opens the first solution after 19 sec but then spends more than two minutes (presumably working on the second question) before he shows the second solution.\\
\textit{Interpretation}: This student seems impatient in the first question of exercises to which he is very fast to show the solutions and prefer to spend more time working with the second questions before he shows the solutions. Maybe he thinks that the first questions are too easy? This behavioral structure could be called \emph{Read-Selectively-Verify}.

\begin{figure}[h!]
\includegraphics[width=\textwidth]{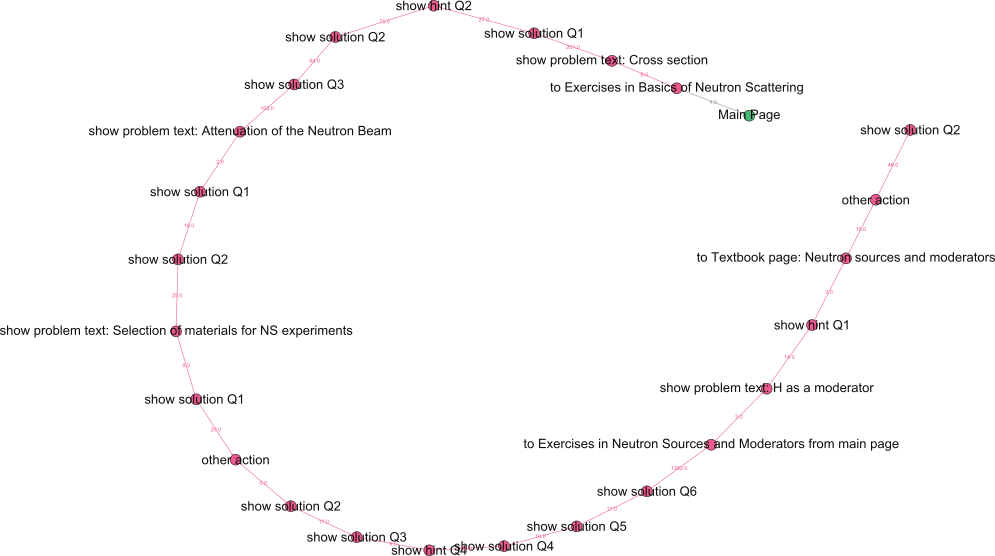}
\caption{Session 618 showing a typical example of a session network in Cluster 6.}
\label{fig:clu06_618}
\end{figure}

\paragraph{Cluster 6} Figure \ref{fig:clu06_618} shows a typical example of a session network from Cluster 6, which has the largest Linear Length within the least complex group (A).\\ 
\textit{Description}: The session has 24 nodes in close to 36 minutes.
In this session the students starts from the main page, clicks after a few seconds on the main page of all exercises and then on a particular exercise after another few seconds. After $\sim 3$ minutes he shows the solution to the first question and after 27 seconds shows a hint to the next question. He then spends just over a minute presumably considering the hint until he opens the solution to the question. He then spends another minute until he shows the solution to the next question. 
After a couple of minutes he clicks on another exercise and immediately shows the solution to the first question and after 18 seconds also shows the solution to the next question. After less than half a minute he clicks a third exercise and after a few (9-20) seconds opens solutions and hints to all of the questions consecutively (there was only one hint in this exercise).
After 23 minutes he goes to a fourth exercise (via an exercise overview page), quickly (14 sec) after shows the first hint, goes to a relevant textbook page, quickly performs an action (search?) and returns to the exercise to show the solution after about a minute.
\textit{Interpretation}: At first glance this student seems to be very active. However after the first exercise the student seems to only glance at the questions, hints and solutions consecutively in a series of exercises and select out only a few specific questions in various exercises to spend around a minute on before opening the solutions. Only in one instance does he spend a longer time before performing the next action (opening another exercise) and it is more likely that he took a coffee break than worked with the previous exercise (and all the open solutions). Only in one instance out of the long consecutive "show everything quickly"-session does he seems to look for information and consider the question again for a minute before opening the solution. \\
This behavior could be explained by a student who printed out all information regarding the problems in the curriculum in order to solve them at another time, but in that case he would probably have spent less time between consecutive clicks. A more likely explanation is that this student shows a surface learning approach since he rarely spends more than a minute working with any question. He may  verify the correct solutions without attempting a full survey-explore-plan-implement cycle of his own.
This behavioral structure could be called \emph{Read-Verify}.

\subsubsection{Example of sessions from group B (Clusters 3 and 7)}
\label{app:examplesB}
\begin{figure}[h!]
\includegraphics[width=0.9\textwidth]{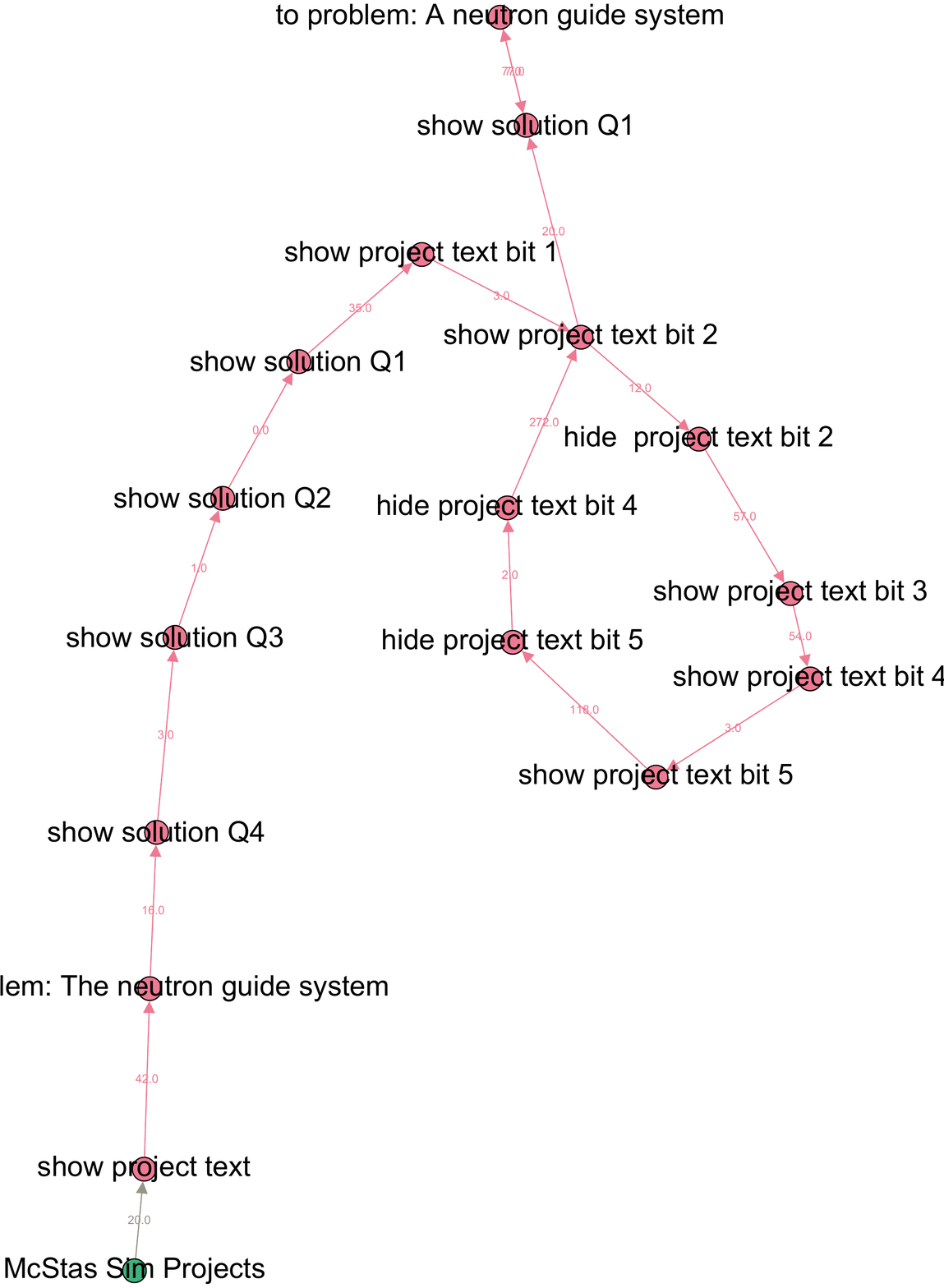}
\caption{Session 813 showing a typical example of a session network in Cluster 3.}
\label{fig:clu03_813}
\end{figure}

\paragraph{Cluster 3} Figure \ref{fig:clu03_813} shows a typical example of a session network from Cluster 3, which scores medium on Complexity and high on Linear Length within that group. \\
\textit{Description}: The session total time is 34 minutes with 17 nodes. 
We see that the student starts on a simulation project page, shows the text of a simulation problem and then goes to a related problem where he quickly opens the solution to all the questions. After half a minute he then opens the first question in the simulation problem, then the corresponding hint and following question of the simulation problem. He spends a minute, before reading the third question and another before opening the fourth and fifth which he then closes again after 2-4 minutes respectively to return to the first part of the project where he shows the hint to the second question after 4.5 minutes. After 20 seconds of revisiting the hint he then navigates several times to a related (simulation) problem where he shows the same solution each time he visits. The first time he shows the solution immediately, next he spends 12 minutes before clicking the solution, the third time he waits 48 seconds and the last time he shows the solution is after close to 9 minutes. \\
\textit{Interpretation}: Since there are no solutions to the simulation project the student is trying to perform, he is in the beginning of the session reviewing the solutions to the related theoretical problem. He then returns to the project and presumably runs the simulation program in the background while comparing his results to the solutions of a related simulation problem in the wiki. This behavioral structure could be called \emph{Embedded-Read-Verify}.

\begin{figure}
\includegraphics[width=0.9\textwidth]{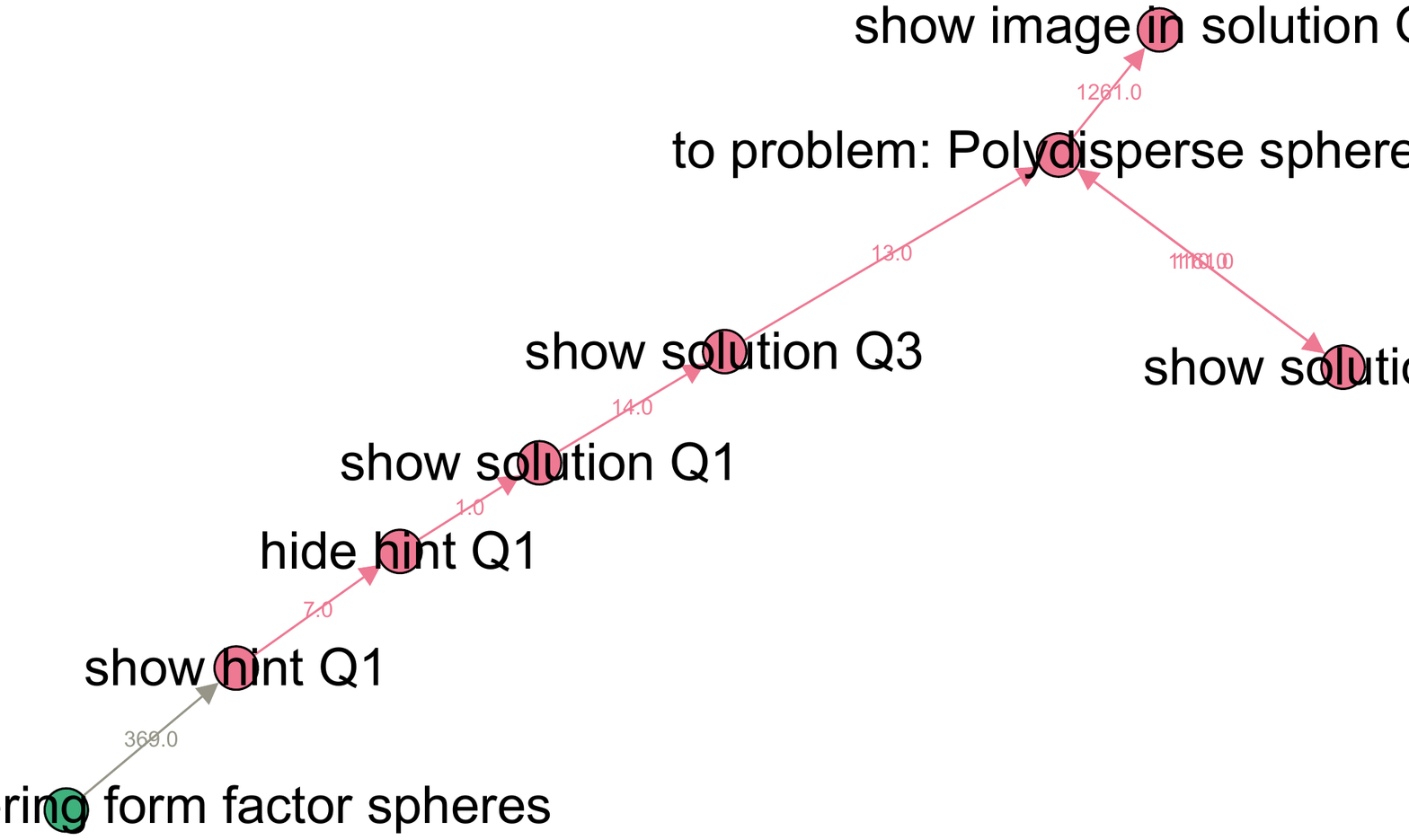}
\caption{Session 1826 showing a typical example of a session network in Cluster 7.}
\label{fig:clu07_1826}
\end{figure}

\paragraph{Cluster 7} Figure \ref{fig:clu07_1826} shows a typical example of a session in Cluster 7, which scores medium on Complexity and low on Linear Length within that group.\\
\textit{Description}: This session has 8 nodes distributed in 49 minutes.\\
The student first spends 6 minutes on a specific problem before opening the first hint and then closing it quickly again to immediately open the solution of the first question which he only glances at (14 sec) before opening the solution to the third and last question. He also only glances at this before going to a second problem where he spends considerable time (19 minutes) before showing the solution to the first question, then spending 2 minutes before clicking on a part of the solution containing a figure which illustrates the solution and after 20 minutes clicking on the figure which gives a more detailed view of the contents.\\
\textit{Interpretation}: The two problems with which the student works are topically related, the second one being a more complex case than the first. In the first question of the first problems the student is to derive a function which he then uses considerable time working on before opening the first hint, and probably sees that he has no further use of the hint so proceeds fast to verifying his solution. In the second and questions of the first problem he probably realizes that he needs a mathematical plotting program to answer the questions and quickly proceeds to the second problem. In the second problem he also needs to derive a mathematical expression which he uses 19 minutes to work on before verifying his solution. He then clicks on part of the solution containing the output from a mathematical plotting program and presumably implements his own function in a plotting program before verifying his results against the ones in the figure upon closer inspection.
In this session the student the student works at length only with the questions not requiring the use of an external tool. He previews the solutions to questions which he needs to use an external tool in order to answer. Only in the end of the session does he implement the external tool and verifies only the most complex result.
This behavioral structure could be called \emph{Embedded-Selectively-Verify}.

\clearpage
\newpage

\subsection{Example of sessions from group C (clusters 4,5,8,9,10)}
\begin{figure}
\includegraphics[width=0.8\textwidth]{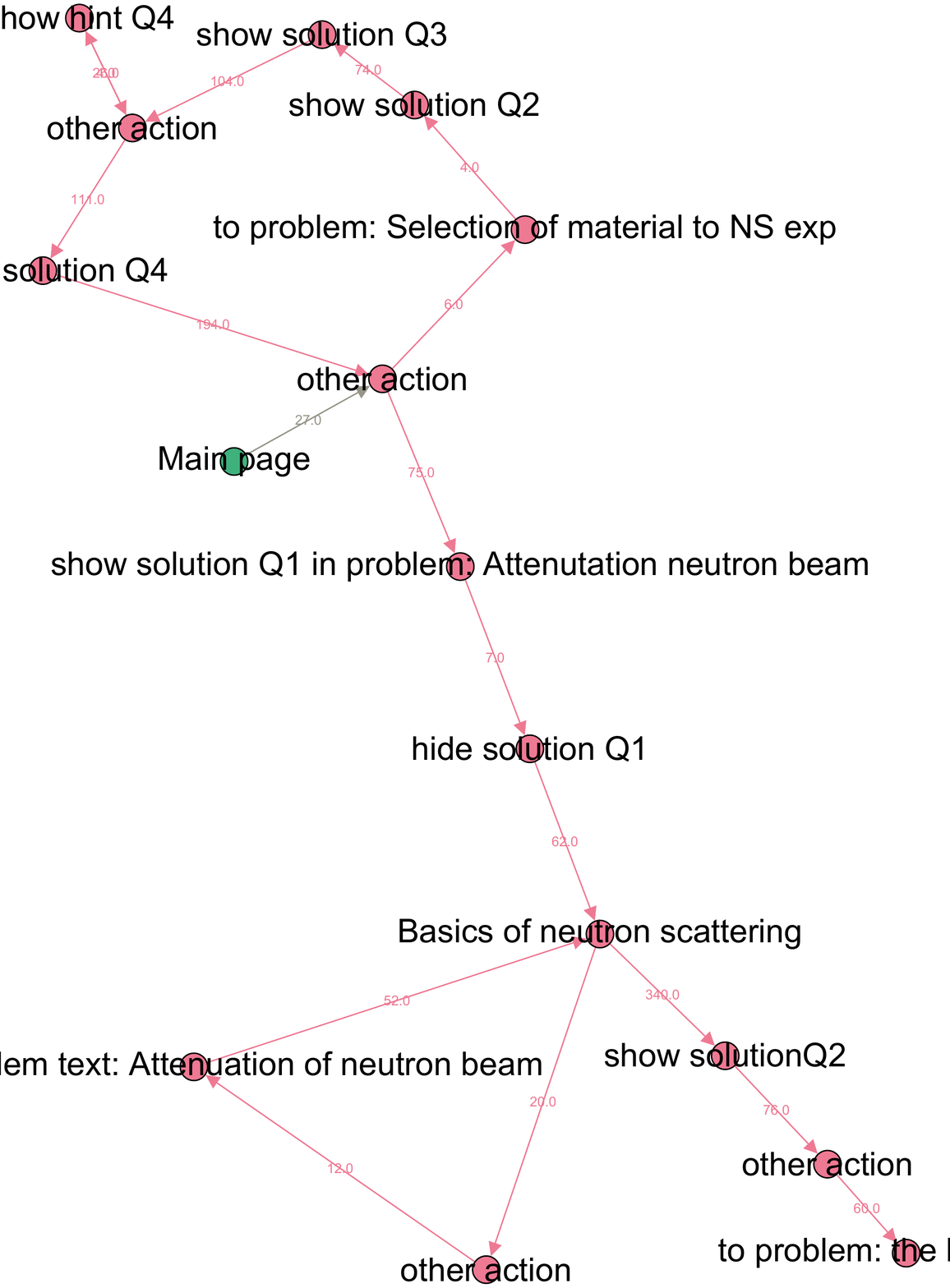}
\caption{Session 1591 showing a typical example of a session network in Cluster 4.}
\label{fig:clu04_1591}
\end{figure}

Figure~\ref{fig:clu04_1591} shows a typical example of a session network in Cluster 4 which scores high on Complexity, Linear Length, and Navigation.\\
\textit{Description}: This session takes 42 minutes and has 16 nodes.
The student navigates a bit on the main problem page and then after roughly half a minute clicks on a specific problem for which he after a few seconds shows the solution to first question and after about a minute shows the solution to the second question. After close to two minutes he navigates on the same page and then quickly thereafter opens the solution to the third question. After half a minute he navigates a bit on the same page and then after almost two minutes opens the solution to the next question. 
After more than 3 minutes he navigates to another problem and after spending a bit more than a minute he shows the solution to the first question but hides it again after glancing at it for 7 seconds. After spending a minute he then navigates to some related text pages where he spends 21 minutes before returning to the description of the problem, navigating a bit on the page for a minute. After close to 6 minutes he shows the solution to the second problem and spends a little more than one minute before proceeding to a third problem which he spends one minute reading before the session ends.\\
\textit{Interpretation}: The student mostly spends at least a few minutes before opening solutions. It seems that in one case he realizes that his answer is wrong after glancing at the solution (quickly hiding it again) and therefore navigates to read information pages for a considerable amount of time before returning to the problem. He then probably works on the second question for some minutes before showing the solution.
The student uses the solutions for verification of his solutions only after spending time on each question and searches for information in cases where his solutions are possibly incorrect. He thus takes advantage of the self-verification and easy access to related learning material features of the wiki-textbook.
This behavioral structure could be called \emph{Read-Verify-Explore}.

\begin{figure}
\includegraphics[width=0.8\textwidth]{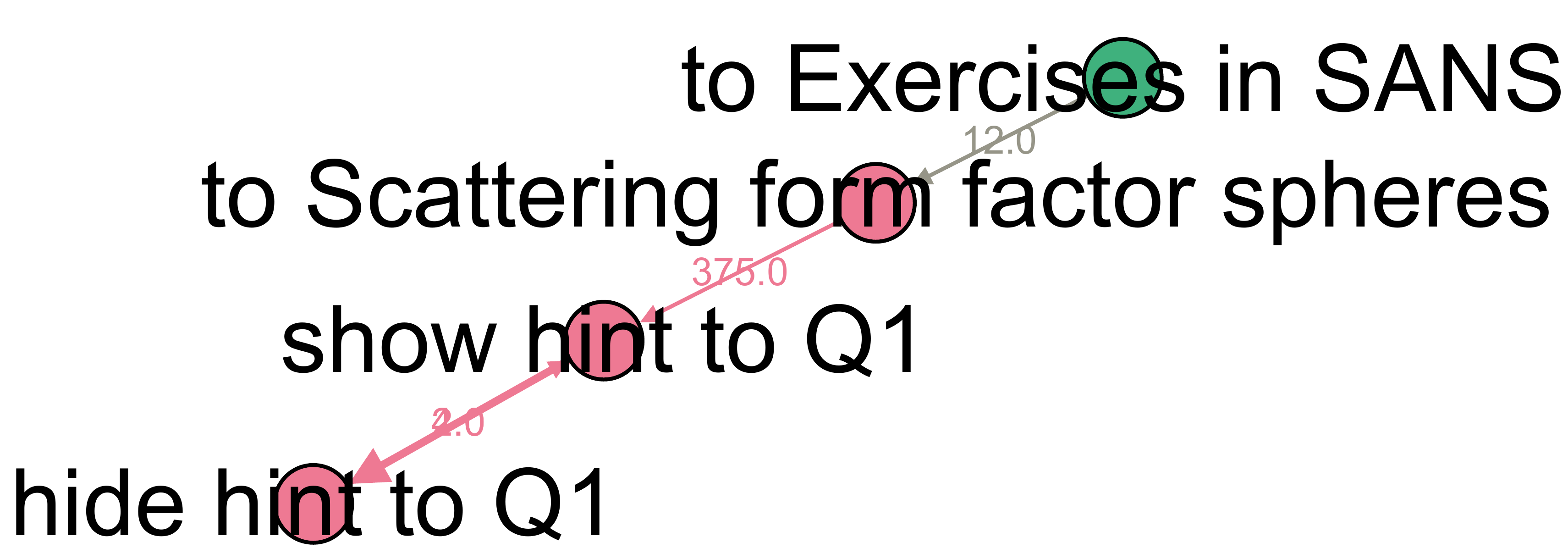}
\caption{Session 118 showing a typical example of a session network in Cluster 5.}
\label{fig:clu05_118}
\end{figure}

\paragraph{Cluster 5} Figure \ref{fig:clu05_118} shows a typical example of a session network in Cluster 5, which scores high on Complexity and low on Linear Length and Navigation.\\
\textit{Description}: This session takes close to 7 minutes and has 4 nodes.
The student navigates first to an overview page of a particular subset of problems and after 12 seconds selects a particular problem. He spends 6 minutes before showing the first hint which he glances for 4 seconds before hiding it again. But after a few seconds he opens the solution once more to glance at it for a few seconds once more and then closes it again.\\
\textit{Interpretation} In this session the student only works with one particular question which he first tries to solve and then several times glances at the solution (show-hide). He does however not work more with his solution between consecutive show-hides so probably he is uncertain about the first verification. This behavioral structure could be called \emph{Read-Peak}.

\begin{figure}
\includegraphics[width=0.8\textwidth]{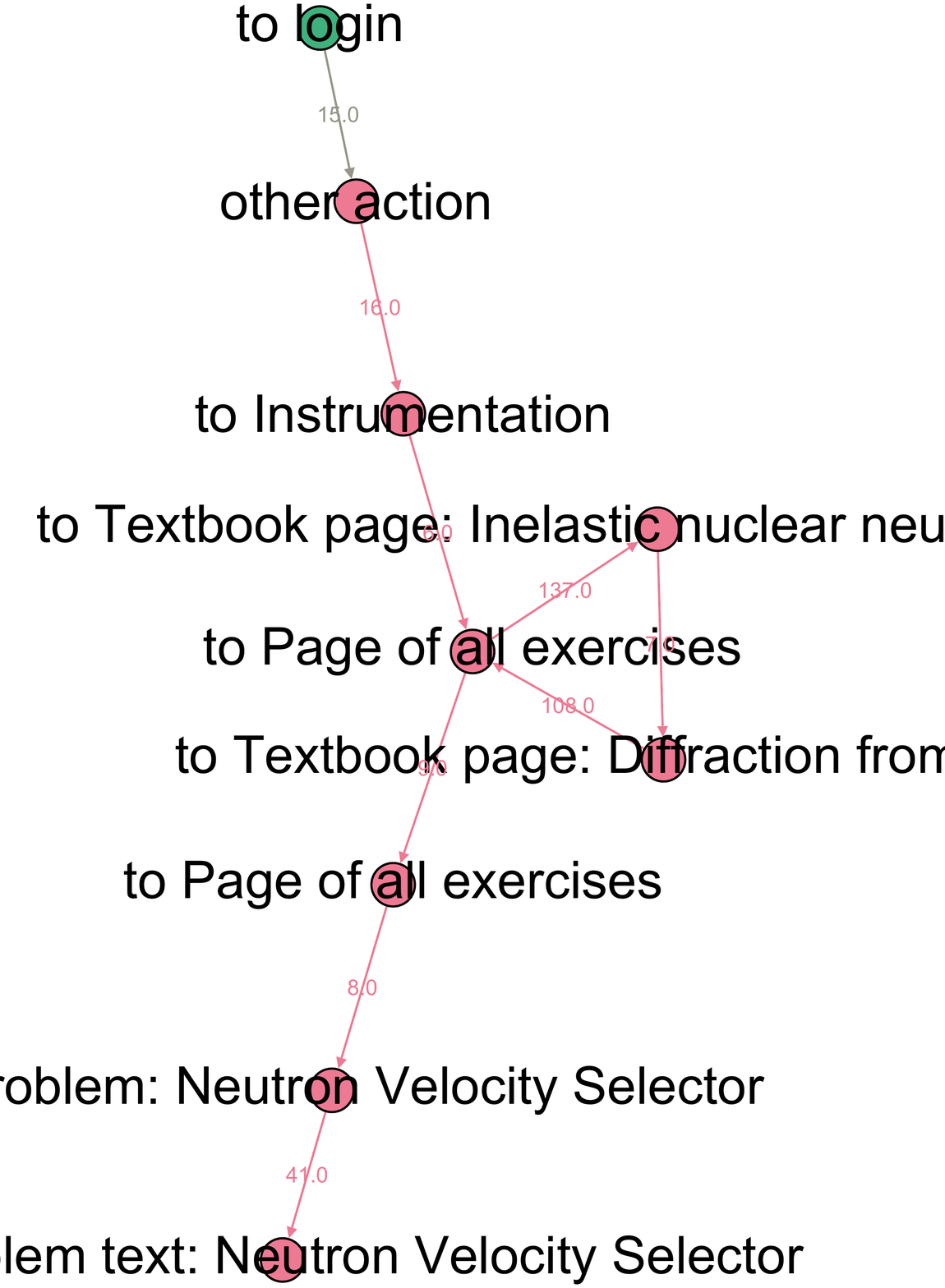}
\caption{Session 1924 showing a typical example of a session network in Cluster 8.}
\label{fig:clu08_1924}
\end{figure}

\paragraph{Cluster 8} Figure \ref{fig:clu08_1924} shows a typical example of a session network in cluster 8, which scores high on Complexity and Navigation, and low on Linear Length within that group. 

\textit{Description}: The session has 9 nodes and lasts for close to 8 minutes.\\
The students starts the session by logging in, navigating to a page with text learning material but spends only about half a minute there in total before proceeding to the overview page of problems where he spends around 3 minutes before proceeding to first one textbook page and then quickly thereafter to another where he spends a couple of minutes before he navigates back to the overview page of problems. He then selects a specific problem and opens the solution after 41 seconds after which the session ends.
\textit{Interpretation}: In this session the student seems to guess about relevant background material for the problem he wants to do in advance of actually engaging with the problem. But after then looking quickly at the problem he may have realized that he did not find the right background information and simply shows the solution. This behavioral structure could be called \emph{Explore}. 

\begin{figure}[h!]
\includegraphics[width=1.2\textwidth]{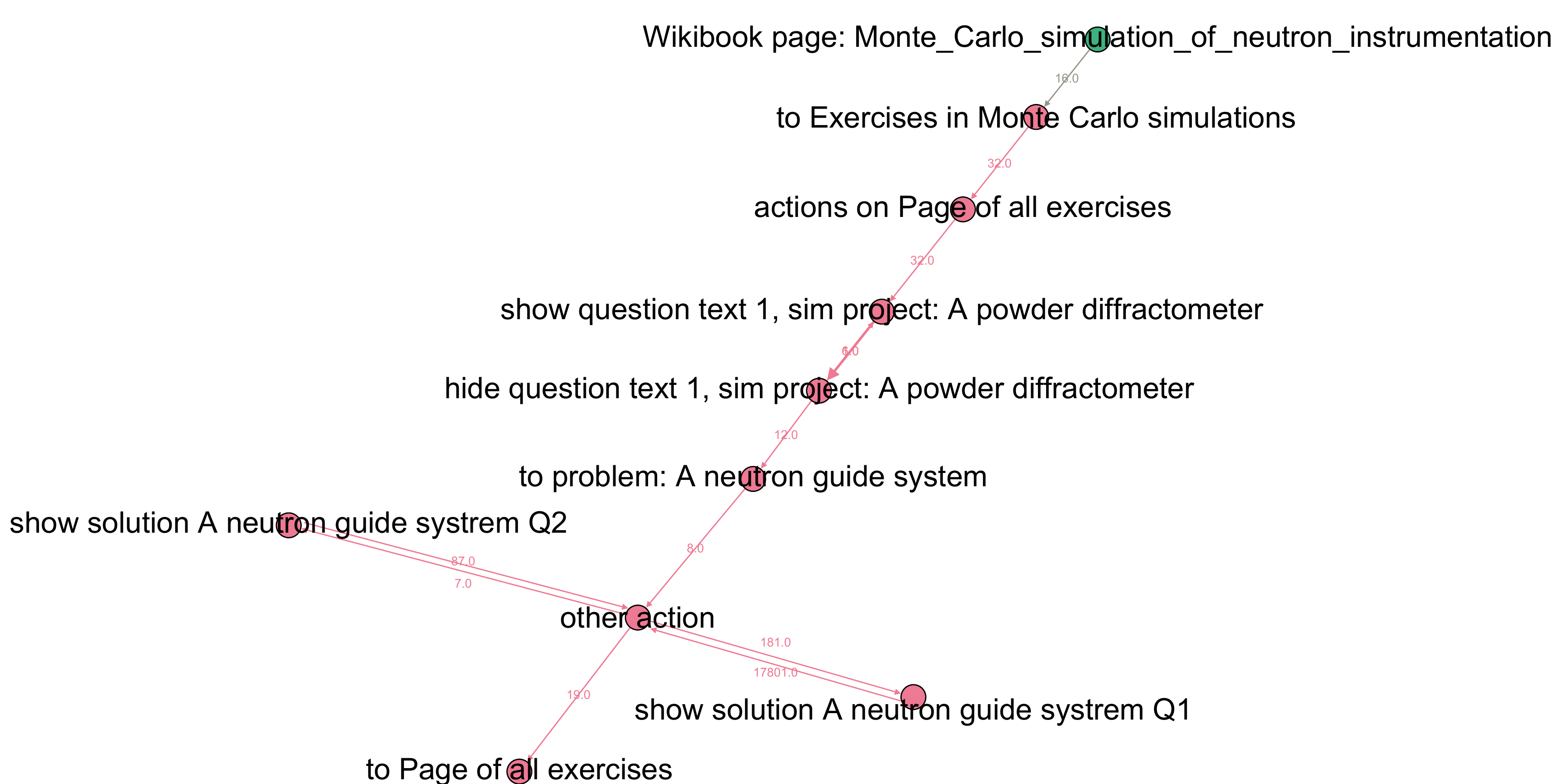}
\caption{Session 1649 showing a typical example of a session network in Cluster 9.}
\label{fig:clu09_1629}
\end{figure}

\paragraph{Cluster 9} Figure~\ref{fig:clu09_1629} shows a typical example of a session network in Cluster 9, which scores high on Complexity and Mutuality and low on Navigation. 
\textit{Description}: The session lasts 332 minutes (close to 6 hours) and has 10 nodes.\\
The students starts at a textbook page with theoretical background for using particular simulation software and after 16 seconds he navigates to the overview of related problems. After half a minute he goes to the page of all problems in order to after another half minute go to a (complex) simulation project page where he shows the first question text. After 6 seconds he hides the text and then shows and hides the text quickly once more. After 12 seconds he then navigates to a simple simulation problem which is highly relevant for the first question of the more complex simulation project. He clicks on a headline and some text in the simple simulation problem during the next 45 minutes but presumably performs the required simulation to answer the question in another program meanwhile. Only after 45 minutes does he show the solution the first question in the simple simulation problem (there is no hint in this case). During the next 15 minutes he seems to be investigating the solution and perhaps trying to solve/simulate the second question (clicking various places on the simple simulation problem page without closing solution to Q1, each click is at least 3 minutes apart). When the 15 minutes have passed he opens the solution to the second question (there is also no hint to this question). After 2 minutes where he probably inspects the solution and click various places on the page he navigates to the overview page of all problems where the session ends.
\textit{Interpretation}: In this session the student seems to go directly from the background info to the relevant simulation project, then realizes that he needs to start with something more simple and then spend substantial time solving a simpler but highly relevant simulation problem without peeking on the solutions before working through the simulation himself. When he actually does open the solutions he appears to inspect them carefully, clicking various places in the page with open solutions several times.
This behavioral structure could be called \emph{Integrated-Interactive}.

\begin{figure}[h!]
\includegraphics[width=0.9\textwidth]{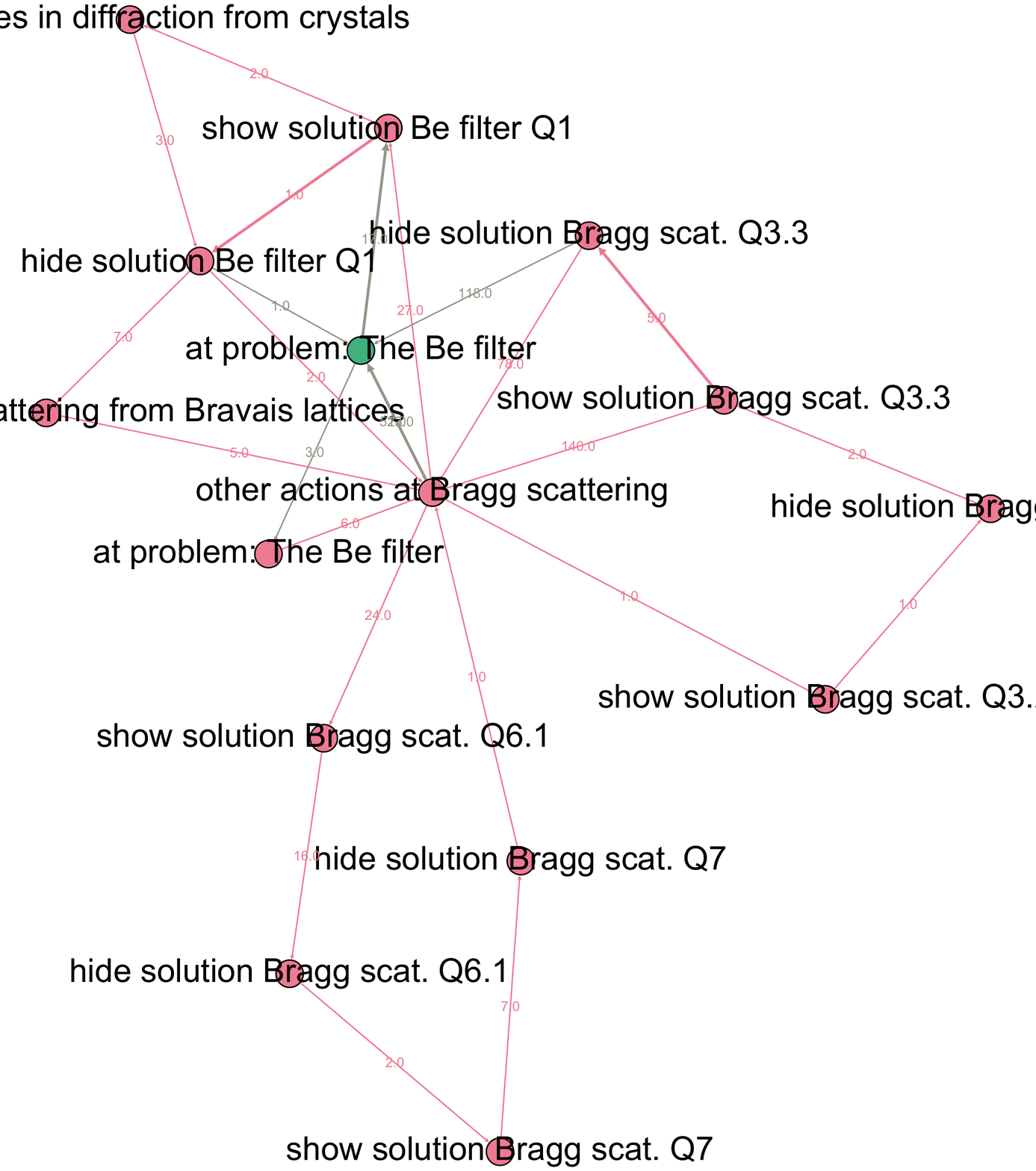}
\caption{Session 1938 showing a typical example of a session network in Cluster 10.}
\label{fig:clu10_1938}
\end{figure}

\paragraph{Cluster 10} Figure \ref{fig:clu10_1938} shows a typical example of a session network in Cluster 10, which scores high on Complexity and Trial \& Error, but shows high variability on the other components. 
\textit{Description}: The session lasts 55 minutes and has 15 nodes.\\
The students starts on a particular problem page which he seems to inspect with 13 clicks on the page (headlines and paragraphs) distributed over 16 minutes in total before he shows the solution to the first question (there is no hint). After glancing at the solution for 17 seconds he opens a related problem (which is posted in the solution). After a few seconds he closes the solution to the first question in the first problem and continues to work on the second problem, clicking a lot of times on various places in the page over the next minute. He the returns to the solution of the first problem which he glances at before closing it again to return to the second problem page where he after half minute opens the solution to the sixth question, inspecting it for 16 seconds before hiding it again and then quickly shows and hides the solution to the next question.
Over the next 1.5 minutes he inspects the page of the second problem (clicks 4 times on paragraphs) before yet again returning to the first problem page which he inspects with 9 clicks over roughly a minute before yet again inspecting the second problem page with 28 clicks in roughly a minute after which he shows the solution to the third question, but hides it within seconds and then the solution to question 4 which he hides just as quickly.
He then returns for a third time to the first problem, this time spending nine minutes inspecting the page in four clicks, before going a fourth time to the second problem where the inspects the page in some clicks and then after 2 minutes and 20 seconds opens again the solution to the fourth question but hides it after a few seconds presumably pondering on this solution for 4 minutes before he inspects this page some more in 7 rapid clicks and then after 43 seconds returning to the first problem for the fourth and final time. Over the next 14.5 minutes he inspects the page, most click are only seconds apart but 2 of them are several minutes apart where he presumably is trying to derive the solution to the original question in the first problem for which he then shows the solution once more but hides again after a few seconds.
\textit{Interpretation}: The student seems to be comparing the content of one problems to the description and solutions of a second problem in the attempt to find a solution to the questions in the first problem. He opens some of the solutions (questions 6 and 7) in the second problems probably because there is one particular keyword figuring both in the first problem and in questions 6 and 7 of the second problem. But after revisiting the first problem he appears to not have found the help he needed and return again to the second problem to open some of the other solutions (questions 3 and 4). He may have got an idea on how to solve the first problem from these solution because he now spends nine minutes at the fist problem page, then he revisits the solution to q4 of the second problem which he uses some minutes to try and solve before revisiting (maybe verifying) the solution. He probably now considers the context of his result by inspecting the page of the second problem because after 4 minutes he inspects this page some more before  finally returning to the first problem and presumably trying to solve it but also inspecting the page. After close to a quarter of an hour here at the first problem page he shows once more the solution but hides it again after only 2 seconds, which may be too fast for a verification of the result and could indicate he is rather giving up.\\
This learner is very persistent in trying to find information that could help him solve one particular problem which is formulated rather openly. He tries to find solutions to relevant problems and probably tries similar approaches to solving his problem. It is not however completely clear if he succeeds and he doesn't seem to spend an adequate time verifying his attempts (hides the solutions quickly). He makes the same steps several times indicating a trial-error approach to problem solving, possibly unsuccessful (in this case).
This behavioral structure could be called \emph{Erratic-Interactive}.

\section{Additional discussion}
\label{app:addDisc}
\subsection{Detailed overview of findings}
\label{app:overviewFindings}

Table~\ref{tab:clusterranking} sums up our characterisation of clusters and groups per component. Clusters have been characterized as uncertain (unc.), low, medium (med.), or high scoring on each of the components of structural behavior. Two non-structural components proved fruitful for characterising clusters and groups. First, we defined a Time Class variable and found that Clusters in Group A all showed a significant over representation of Short length (5-25 minute) sessions, while Cluster 3 showed an over representation of Middle length (25 minutes- 1.4 hours) sessions, and Cluster 4 showed an over representation of Long (1.4hours to 3 hours) and Extensive (3 hours +) sessions. Second, we define the $\mu$-parameter which measures the degree to which showed hints and solutions are hidden again during a session. 

\begin{landscape}
\begin{table}
\caption{A summarizing description and ranking of each cluster characteristics.}

\begin{tabular}{p{1cm} c c c p{1.5cm} c c p{1.25cm} p{2cm} p{4cm}}
\hline
Group & $\mean{\mu}$ & Cluster & Complexity & Linear Length & Navigation  & Mutuality & Trial\& Error& Time Class& Extracted behavioral structure   \\ 
\hline
\multirow{3}{*}{A} & \multirow{3}{*}{high} &1 & low & med. & med. & med. & low & Short &  Read-Selectively-Verify\\
& & 2 & low & low & med. & med. & low & Short& Read
\\
& & 6 & low & high & med. & low & low & Short &  Read-Verify\\
\hline
\multirow{2}{*}{B} &\multirow{2}{*}{unc.} & 3 & med. & high & med.  & med. & low & Middle &Embedded-Read-Verify \\
& & 7 & med. & low & med. & med. & low & - & Embedded-Selectively-Verify\\
\hline
& & 4 & high & high & high & unc. & low & Long/ Ext. &Read-Verify-Explore\\
\multirow{4}{*}{C} & \multirow{4}{*}{low} & 5 & high & low & low & med. &  low& - & Read-Peak\\
& & 8 & high & low & high  & low & low &  - & Explore\\
& & 9 & high & med & low & high & low & - & Integrated-Interactive\\
& & 10 & high & unc. & unc. & unc. & high& -  & Erratic-Interactive \\
\hline
\end{tabular}
\label{tab:clusterranking}
\end{table}
\end{landscape}

\subsection{Additional discussion: The wiki-textbook as learning material}
\label{app:wikiLearning}
The wiki-textbook format is meant to address some of the challenges that a normal textbook format faces. For instance, knowledge represented in the wiki-textbook can be easily updated and can also be crowd-sourced to students as part of their learning inspired by the literature on using wikis as learning tools \citep{augar2004,parker2007,lin2009,matthew2009,karasavvidis2008}. By being web-based it is accessible through multiple channels (and can be printed), and can be searched and navigated in different ways as compared with a normal textbook. In this study, we have seen evidence of students making use of the functionalities specific to a wiki-textbook. Group C in particular seems to represent different ways of using the wiki-textbook problems, which would seem difficult to employ with other textbook formats. 

For both researchers and teachers the fact that data about student use is collected in server logs may add to the affordances of wiki-textbooks as a teaching and learning tool. We saw that for the very linear sessions in Group A hints and solutions were shown shortly after entering the page and often abandoned before some time. Combined with the fact that, short sessions were over represented, this may mean that there is a risk that when students engage only shortly with wiki-textbook problems they may be employing surface strategies, where they read solutions prematurely. In a blended environment, teachers may then engage productively with students who finish too soon. For researchers, these kinds of behaviors could help when choosing appropriate situations to investigate further, for example, by interviewing students. 

Clearly, there are many ways for students to use the wiki-textbook to learn in a university physics course. We have identified 10 behavioral structures. As we will discuss below, these behavioral structures should be coupled with other observations and perhaps even intentions of the students to understand how on-line teaching materials such as the wiki-textbook may support or hinder learning. However, the results from each group do give hints about which behavioral structures may be linked to more beneficial learning strategies. For example, many sessions in Cluster 4 are long, and hints/solutions tend to be hidden, when they have been shown. Some of this may stem from students navigating the web-site during the problem solving session. This may indicate that they have been motivated to pursue knowledge that they are missing to solve the problem. This autonomous searching for pieces that may help one construct a solution would be consistent with a deep learning strategy. Thus, we would expect students to learn more, when they employ such a strategy than if they only read a solution. The existence of such a strategy might also be considered an affordance of the wiki-textbook. It is not seen or not seen to near the same extent in more linear media, such as a standard textbook or an e-book. On the other hand, we have also identified the \emph{Erratic-Interactive} behavioral structure, which may indicate students who are over-using the functionality of the wiki-textbook. 

An interesting feature that we have found is that it seems that some behavioral structures consist of compositions of behavioral structures. For instance, from Cluster 3 we identified the \emph{Embedded-Read-Verify} behavioral structure. Such structures might be linked with a more strategic approach to learning. For example, it may be that a student needs the solution to one problem in order to solve another more complex problem. Then, the student may employ a very linear behavioral structure for part of the problem solving session. While the student may not learn much about the solution to the embedded problem, the student may learn a great deal about how that solution fits into a broader scheme. 

Just as students need to learn how to  use a normal textbook, and indeed any technology for learning, students will likely need to learn how to use a wiki-textbook to best suit their needs in a given situation. 


\end{document}